\documentclass[twocolumn,10pt]{IEEEtran}
\usepackage{algorithm,algorithmic,amsbsy,amsmath,amssymb,epsfig,bbm,mathrsfs,multirow,amsthm}
\usepackage[T1]{fontenc}
\usepackage[latin9]{inputenc}
\usepackage{amsmath}
\usepackage{fixltx2e}
\usepackage{algorithm}
\usepackage{array,multirow,graphicx}
\usepackage[hyphens]{url}

\DeclareMathAlphabet{\mathpzc}{OT1}{pzc}{m}{it}

\newtheorem{proposition}{Proposition} 
\newtheorem{assumption} {Assumption}

\begin{document}

\title{Charging and Discharging of Plug-In Electric Vehicles (PEVs) in Vehicle-to-Grid (V2G) Systems: A Cyber Insurance-Based Model}
\author{Dinh Thai Hoang$^1$, Ping Wang$^1$, Dusit Niyato$^1$, and Ekram Hossain$^2$	
\\$^1$ School of Computer Science and Engineering, Nanyang Technological University, Singapore
\\$^2$ Department of Electrical and Computer Engineering, Universtity of Manitoba, Canada \vspace{-5mm}}

\maketitle

\begin{abstract}
In addition to being environment-friendly, vehicle-to-grid (V2G) systems can help the plug-in electric vehicle (PEV) users in reducing their energy costs and can also help stabilizing energy demand in the power grid. In V2G systems, since the PEV users need to obtain system information (e.g., locations of charging/discharging stations, current load and supply of the power grid) to achieve the best  charging and discharging performance, data communication plays a crucial role. However, since the PEV users are highly mobile, information from V2G systems is not always available for many reasons, e.g., wireless link failures and cyber attacks. Therefore, in this paper, we introduce a novel concept using cyber insurance to ``transfer'' cyber risks, e.g., unavailable information, of a PEV user to a third party, e.g., a cyber insurance company. Under the insurance coverage, even without information about V2G systems, a PEV user is always guaranteed the best price for charging/discharging. In particular, we formulate the optimal energy cost problem for the PEV user by adopting a Markov decision process framework. We then propose a learning algorithm to help the PEV user make optimal decisions, e.g., to charge or discharge and to buy or not to buy insurance, in an online fashion. Through simulations, we show that cyber insurance is an efficient solution not only in dealing with cyber risks, but also in maximizing revenue of the PEV user. 
\end{abstract}


\section{Introduction}
\label{sec:introduction}

\begin{figure*}[h]
	\begin{center}
		$\begin{array}{c} \epsfxsize=4.0 in \epsffile{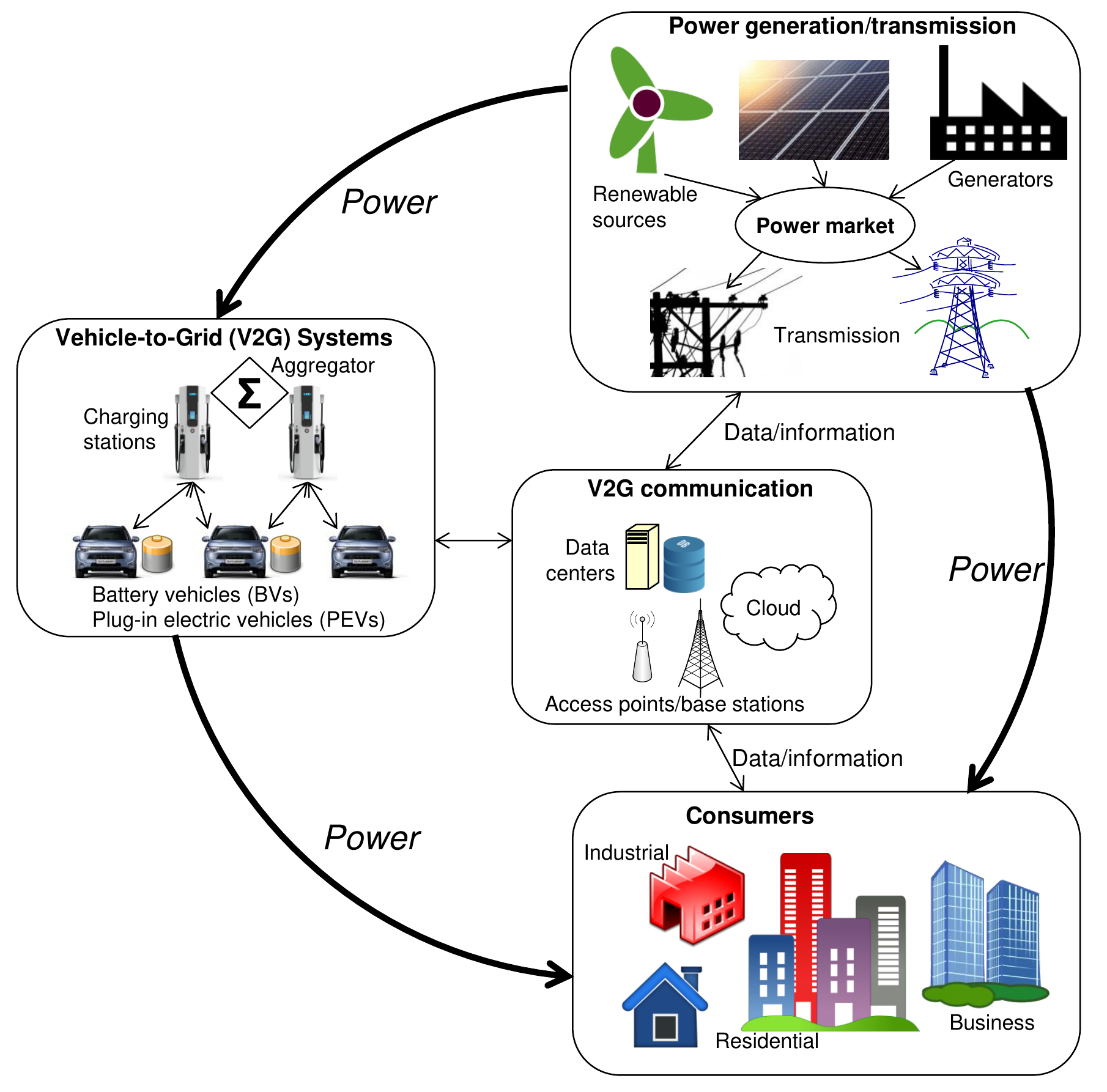} \\ [-0.2cm]
		\end{array}$
		\caption{V2G architecture.} 
		\label{fig:V2G _Architecture}
	\end{center}
\end{figure*}

One challenge of the current power grid is to provide sufficient capacity and cost-effective energy storage. The energy storage is used as a tool by the power grid operator to efficiently manage the generation and transmission of the electricity, i.e., supply and delivery, to meet dynamic and unpredictable consumer demand. A traditional approach is to deploy large generators which can be relatively ineffective due to its long delay response (minutes) and can cause underutilization (spare capacity). In smart grid, ancillary services such as load regulation, spinning reserve, non-spinning reserve, and replacement reserve to support the continuous flow of electricity have been used to alleviate this problem. However, the introduction of renewable sources, the energy supply of which depends on natural conditions, aggravates the problem due to the fluctuating and unpredictable characteristics. Therefore, the vehicle-to-grid (V2G) systems have been considered as a promising solution. In V2G systems, battery vehicles (BVs) or plug-in electric vehicles (PEVs) can be used as energy storage devices. Although their battery capacity is limited, they are suitable for short-time ancillary services given their small response time as well as lower standby and capital costs. 

The effectiveness of V2G systems depends on the number of PEVs participated and how good the data, e.g., information of PEVs and charging stations, is exchanged between V2G operator and PEVs in order to optimize system operations. For example, the V2G operator can economically manage its generators if the amount of energy reserved from V2G systems can be accurately estimated. Likewise, the PEVs can choose to charge or discharge their batteries to maximize the performance and minimize the cost. However, since PEVs are mobile vehicles and the information about V2G systems is transmitted to the PEVs through wireless links, the V2G data communication is unreliable and vulnerable to cyber attacks which can violate confidentiality, authenticity, integrity, and availability requirements of the data exchange in V2G systems. A number of cyber risks have emerged to the V2G systems. The majority of research works focus on mitigating the risks by protecting the systems and preventing adverse effects from the attacks. However, it is well known that no single solution can completely avoid the risks and their damage. Recently, \emph{cyber insurance} has been introduced as an efficient solution to alleviate damages for cyber customers. With cyber insurance, PEVs' risks are ``transferred'' to a third party~\cite{Gordon2003AFrame}, thus PEVs are protected from cyber attacks and compensated for their losses if they are victimized to such attacks.

In this paper, we introduce a novel idea of using cyber insurance for PEVs in V2G systems. First, we present an overview of V2G systems, including data communication and cyber risks. Some related works on V2G system security are also reviewed. We then introduce a short survey about cyber insurance. This survey is used to provide basic concepts as well as fundamental knowledge about cyber insurance. Finally, we propose a novel model using cyber insurance for a PEV user in a V2G system. Specifically, we use a Markov decision process framework to formulate the energy cost optimization problem with the aim of minimizing the average total energy cost for the PEV user. In addition, we also propose a learning algorithm to help the PEV user make optimal decisions, i.e., charge or discharge and buy or not to buy insurance, given its current state, e.g., battery level and insurance status, in an online fashion. The proposed solution not only minimizes the average total cost for the PEV user, but also maximizes the PEV's revenue without a need of PEV's prior knowledge on the risk, e.g., the probability of information unavailability. The proof of the convergence for the proposed learning algorithm is also provided in this paper. Through simulations, we demonstrate that adopting a cyber insurance model can provide an efficient  solution to the cost minimization problem for the PEVs. 

The rest of the paper is organized as follows. In Section~\ref{sec:Overview_V2G}, an overview of V2G systems and their security problems are presented. Section~\ref{sec:O_CI} provides basic concepts and fundamental knowledge about cyber insurance. Then, we introduce the idea of using cyber insurance to mitigate the risks and propose the learning algorithm to minimize the cost for the PEV user in Section~\ref{sec:RM_CI_PEV}. Finally, future research directions are highlighted in Section~\ref{sec:Future Reserach}, and the conclusions are presented in Section~\ref{sec:conclusion}.
\section{Overview of V2G Systems}
\label{sec:Overview_V2G}

\subsection{Vehicle-to-Grid (V2G) Systems}

\subsubsection{Introduction}

Vehicle-to-grid (V2G) describes a system in which plug-in electric vehicles (PEVs), e.g., electric cars and plug-in hybrids, communicate with the power grid to facilitate demand response services by either charging or discharging energy. On one hand, if the PEVs perform charging from the power grid, the energy will be stored in their batteries for traveling and storing. On the other hand, if the PEVs perform discharging to the power grid, the energy from their batteries will be returned to the power grid with the purpose of stabilizing energy demand~\cite{Wei2014GT-CFS}. For example, when the energy supply from generators exceeds demand, e.g., during off-peak hours, a low energy price can be offered to incentivize PEVs to charge their batteries from charging stations~\cite{Tan2016Pareto}. By contrast, when the energy supply cannot meet the demand, e.g., during peak hours, PEVs can sell their energy back to the power grid. Hence, PEVs can act as an energy reserve. As such, PEVs are expected to potentially offer unprecedented benefits to the grid. For example, it is estimated that ancillary services of PEVs account for 5-10\% of electrical cost, or about \$12 billion per year in the U.S.~\cite{Kempton2005Vehicle}.

\begin{table*}[t]
	\caption{Advantages and disadvantages of centralized and decentralized control solutions} 
	\centering 
	\begin{tabular}{|l|l|l|}
		\hline
		&\textbf{Advantages}&\textbf{Disadvantages}\\
		\hline
		\hline
		\textbf{Centralized solution} & $\bullet$ Maximize revenue for the provider and PEVs& $\bullet$ Complex and expensive communication infrastructure \\
									& $\bullet$ Control energy and ancillary services efficiently & $\bullet$ Require a powerful central controller and a backup data storage \\
									& &	$\bullet$ Require full information from the PEVs \\
									& &	$\bullet$ Decisions of the PEVs are controlled by the provider \\
									& &	$\bullet$ Must be able to handle a large amount of data at the same time \\
									& &	$\bullet$ Privacy of the PEVs can be vulnerable \\
									& &	$\bullet$ Can be delayed or interrupted due to the system overload  \\
									& & or cyber attacks \\
		\hline		
		\textbf{Decentralized solution} & $\bullet$ Able to adapt to a large number of PEVs  & $\bullet$ Require efficient decentralized control solutions for the PEVs \\
										& $\bullet$ Less communication and infrastructure required  & $\bullet$ Require methods to predict demands of the PEVs for the provider \\
										& $\bullet$ Fast and convenient services since decisions &  $\bullet$ The PEVs must find approaches to protect themselves   \\
										& are made and controlled by the PEVs   &  from cyber attacks \\
										& $\bullet$ Preserve individual authority  &  \\
										& $\bullet$ Better fault tolerance   &  \\
		\hline
	\end{tabular}
	\label{tab:smart_control}
\end{table*}

\subsubsection{Architecture}

Fig.~\ref{fig:V2G _Architecture} shows a general architecture of a V2G system with interactions among power generation/transmission, power consumers, and PEV users~\cite{Guille2009A conceptual}. The power systems include convention generators, renewable sources, and transmission facility. The power systems supply energy to both consumers (e.g., residential, industrial, and business) and V2G systems. The V2G systems are composed of PEVs connected with the power grid through public and private charging stations and aggregators. An aggregator is a mediator controlling and optimizing energy flow between power grid and V2G systems. The V2G systems act as both energy storage and consumers. V2G communication provides data and information exchange among power systems, power consumers, and V2G systems, and it consists of communication infrastructure (e.g., wireless networks) and processing facilities (e.g., cloud computing and data center). With the V2G communication infrastructure, the power system operators can collect necessary data from V2G systems and consumers, then optimize power generation and ancillary services from PEVs efficiently.

PEV users can make a long-term agreement/contract with the V2G operator to make charging and discharging more predictable. For example, the operator can offer battery maintenance service in exchange for PEV users agreeing to charge and discharge the battery to meet the requirements of the V2G systems. With this approach, centralized control of charging and discharging process can be implemented to achieve the maximum efficiency. However, to achieve such a goal, status monitoring and information update are necessary for V2G systems. The V2G systems should be able to obtain the timely conditions of both moving and parking PEVs. The conditions can be PEVs' locations, battery capacities, battery state-of-charge, expected time to arrive at and leave charging stations. Using this information, the V2G system can estimate the amount of energy to charge and to receive from PEVs in certain areas.

Alternatively, some PEVs can participate in V2G systems voluntarily without making long-term commitment with the V2G operator. For example, the operator can offer different incentives for charging and discharging energy by PEVs depending on current load and supply of the power grid. The PEV user individually considers the current location, i.e., charging stations' locations, the battery state-of-charge, and energy price to decide to charge (or discharge) its battery or not. With this approach, charging and discharging decisions are made by PEV users in a distributed fashion. Therefore, the V2G systems must provide information about the incentive to motivate the users in such a way that the system efficiency is maximized.

\subsubsection{Smart charging/discharging control}

As the number of PEVs increases, implementing smart charging/discharging control solutions has become more and more important to avoid large expenditures and negative impacts on the power gird. In general, charging/discharging control is classified into two groups, i.e., centralized and decentralized solutions~\cite{Villalobos2014Plug}. For the centralized solution, all charging/discharging processes of PEVs will be controlled by an authorized energy service provider. By contrast, for the decentralized solution, charging/discharging decisions are made and performed by the PEVs themselves. Each solution has its own advantages as well as disadvantages as shown in Table~\ref{tab:smart_control}. Although the centralized approach can achieve optimal performance more easily for both the provider and PEVs, it may not be practical to implement as the PEVs cannot control their charging/discharging processes by themselves. Therefore, in actual systems, the decentralized solution is more preferable~\cite{Villalobos2014Plug}. 

\begin{table*}[t]
	\caption{Battery capacity and technologies} 
	\centering 
	\begin{tabular}{|c|c|c|c|c|}
		\hline
		Car model/EV type & Battery & Range & Charging time \\
		\hline
		\hline
		Chevrolet Volt & 16kWh, Li-manganese/NMC, liquid cooled, 181kg  & 64km & 10h at 115V AC, 15A	\\
		& & &  4h at 230V AC, 15A 	\\
		\hline
		Toyota Prius & 3 Li-ion packs, one for hybrid, two for EV, 50kg & 20km&  3h at 115V AC 15A	\\
		& & &  1.5h at 230V AC 15A 	\\
		\hline
		Mitsubishi iMiEV  & 16kWh; 88 cells, 4-cell modules; Li-ion; 150kg; 330V & 128km  & 13h at 115V AC 15A 	\\
		& & &   7h at 230V AC 15A 	\\
		\hline
		Nissan Leaf  & 30kWh; Li-manganese, 192 cells; air cooled; 272kg  & 250km & 8h at 230V AC, 15A 	\\
		& & &   4h at 230V AC, 30A 	\\
		\hline
		Tesla S  & 70 and 90kWh, 18650 NCA cells of 3.4Ah;  & 424km  & 9h with 10kW charger; 	\\
		& liquid cooled; 90kWh pack has 7,616 cells;  540kg &  &   120kW Supercharger,  80\% charge in 30 min	\\
		\hline
		BMW i3  & 22kWh (18.8kWh usable), LMO/NMC,  & 130-160km & 4h at 230VAC, 30A; 	\\
		& large 60A prismatic cells,  204kg  & &   50kW Supercharger; 80\% in 30 min 	\\
		\hline
		Smart Fortwo ED  & 16.5kWh; 18650 Li-ion  & 136km & 8h at 115VAC, 15A; 	\\
		& & &   3.5h at 230VAC, 15A 	\\
		\hline
	\end{tabular}
	\label{tab:Battery capacity}
\end{table*}

\subsubsection{Benefits}

V2G systems offer many benefits to the power grid and also PEV users~\cite{Yang2011P2}. 
\begin{itemize}
\item \emph{Diminishing environmental pollution:} Different from conventional vehicles using fossil fuel, PEVs can diminish significantly environmental pollution even when considering power generation emissions. It is estimated that by replacing a conventional car by a PEV, CO$_2$ emissions can be dropped by 2.2 tons per year~\cite{Islam2014Integrating}. 
\item \emph{Enhancing ancillary services:} In practice, there are many cars traveling on the road for only 4-5\% of the day, while they spend the rest of time for parking. This implies that we can utilize such electric vehicles to facilitate the ancillary services in V2G systems, e.g., spinning reserves, reactive power support, frequency and voltage regulation, to balance supply and demand for reactive power. These services can be used to reduce an overall cost of V2G systems, thereby decreasing energy prices for customers and improving load factors.  
\item \emph{Improving quality of services for PEV users:} Due to the development of battery technologies, V2G systems enable very fast energy supply response time in which the charging and discharging responses can be performed in milliseconds. Furthermore, there is no significant running cost of the unit commitment operations. Therefore, quality of services for PEV users, e.g., serving time, can be improved considerably. 
\item \emph{Supporting renewable energy:} The power quality from renewable sources such as solar and wind generators can be greatly improved by using PEVs as storage and filter devices. The combination of PEVs and renewable energy sources can make the power grid more stable and reliable.
\item \emph{Rising revenue to PEV users:} PEV users can receive monetary reward for discharging energy or other support benefits from V2G operators in participating in the system. Thus, by adopting intelligent energy management solutions, the PEV users can balance their demands and charging/discharging processes, e.g., charging during non-peak hours and discharging during peak hours, to obtain more revenues. 
\end{itemize}

\begin{table*}[!]
	\caption{Wireless communication technologies in V2G systems} 
	\centering 
	\begin{tabular}{|c|c|c|c|c|c|}
		\hline
		Technology & Operating frequency & Covered distance & Advantage & Disadvantage & Ref.	\\
		\hline
		\hline
		& 868 MHz (Europe) &   & Easy to deploy, 	& High interference, weak security	&	\\
		ZigBee 	& 915 MHz (North America) & 10-100 m & require low bandwidth, 	& short range communication,	&	\cite{Lam2011ZigBee} 	\\
				& 2.4 GHz (Worldwide) & 	& low power consumption 	& high delays	&	\\
		\hline
		Near Field &  & & Convenience, versatility,	& Very short range communication, &	 \\
		Communication  & 13.56 MHz & 5-10 cm & safer than credit cards & lack of security, expensive &	\cite{Steffen2010Near}	\\
		\hline
		 &  &   & Widely used, feature simplicity, 	 & Only connect two devices at once,   &		\\
		Bluetooth & 2.4 GHz & 1-100 m & low power requirement,   & short range communication, & \cite{Conti2011B4V2G} \\
		& & & low interference  &   weak security & \\
		\hline
		 &  &   & Popular standard for V2G systems,	& Unable to modify and difficult	&	\\
		IEEE 802.11p & 5.85-5.925 GHz & 500-1000 m & suitable for high-speed vehicles	&  to handle 	a large number & \cite{Anbagi2016WAVE}	\\
		 & & & and QoS-required applications,	&	of users,  no authentication	& \cite{Msadaa2010A comparative}	\\
		& & & high data transfer speed	&  prior to data exchange	&	\\
		\hline
		&  &  &	 Similar features as IEEE 802.11p, 	& Expensive implementation cost,	&	\\
		WiMAX  & 2-6 GHz & 2-5 km  & but longer range communication, & high power consumption, 	&	\cite{Msadaa2010A comparative}	\\
		& & & and higher data transfer speed &	vulnerable by jamming attack  & \cite{Jatav2014WiMAX}	\\
		& & & &	and eavesdropping  &	\\
		\hline
	\end{tabular}
	\label{tab:Wireless communication}
\end{table*}

\subsubsection{Electric vehicle battery}

Different from conventional batteries used in electronic devices such as mobile phones and laptops, batteries for electric vehicles must be designed to prolong the running time with high power (up to a hundred kW) and high energy capacity (up to tens of kWh). In addition, these batteries should have a limited space and weight.  Extensive research efforts are exerted worldwide to invent new advanced vehicle battery techniques which are more suitable for PEVs. In Table~\ref{tab:Battery capacity}, we summarize the advanced vehicle battery technologies which are currently implemented in the real world~\cite{Battery_Type}. In Table~\ref{tab:Battery capacity}, it can be observed that batteries with heavy weights usually offer longer traveling time. However, if the battery is heavy, it will cause inefficient performance for PEVs because the heavy battery will limit the PEVs' speed and consume more energy to carry. Therefore, the balance between the performance and weight of the battery needs to be considered for the future development of electric vehicle batteries.

\subsubsection{Data communications}

In V2G systems, data communication between PEVs and V2G infrastructure is the most crucial step to achieve the best performance for both PEV users and V2G system operators because the operators need information about PEVs' demands to control the energy resources distributed over large geographical areas, meanwhile PEV users need V2G infrastructure information to optimize their energy costs. In this case, wireless communication is the best solution for V2G applications for many reasons. 
\begin{itemize}
\item \emph{Mobility:} PEVs are mobile vehicles, hence wireless communications are the best choice because V2G systems cannot use wires to connect to PEVs. 
\item \emph{Fast and convenient:} Data exchanged between PEVs and V2G infrastructure is often small in size and intermittent over time. So, by using wireless communications, the information will be updated timely and quickly. 
\item \emph{Efficiency with low cost:} Wireless communications allow data to be transmitted to multiple PEVs simultaneously in a wide area coverage. 
\end{itemize}

In Table~\ref{tab:Wireless communication}, we list different wireless communications technologies which have been implemented and developed for V2G systems. From Table~\ref{tab:Wireless communication}, it is observed that each wireless communication technology has its own advantages as well as disadvantages, and it is suitable for PEVs in specific cases. For example, for a short-range data communication, e.g., between a PEV and a charging station when the PEV is charging at that station, ZigBee protocol can be adopted since it consumes less energy for data communications. However, for a long-range data communication, IEEE 802.11p and WiMAX technologies should be used as they are standard protocols for communication over long distances in V2G systems.

\subsection{Security Requirements and Cyber Risks in V2G Systems}

Although wireless technologies bring many advantages, they also raise some security issues for V2G systems. Therefore, the cyber security for data communications between PEVs and V2G infrastructure should be assured in order to protect the smart grid from the cyber attacks such as price tampering and system congestions by malicious software. In this section, we discuss security requirements and some potential approaches to deal with cyber attacks in V2G systems.

\subsubsection{Security requirements}

V2G systems possess the following cyber security requirements.

\begin{itemize}
	\item \emph{Confidentiality:} Data exchanged between PEV users and the V2G operator must be kept confidential. The identity of PEVs users as well as their interaction, i.e., charging and discharging, with the operator must be maintained privately. The cyber attacks to the confidentiality of V2G systems can cause business disadvantages to the V2G operator if its competitor has important information about system operations, e.g., energy price offered to PEV users.
	\item \emph{Authenticity:} The identities of PEVs and the operator must be assured before and during data communications. The operator may miscalculate the V2G system capacity if the identity of PEVs is falsely authenticated. Authentication methods taking specific requirements of V2G systems into account have to be developed. For example, the authentication should be customized and optimized for PEVs [6].
	\item \emph{Integrity:} The integrity ensures that the data exchanged between PEVs and operator will not be modified by attackers. The maliciously modified data such as the number of online PEVs, battery capacity and state-of-charge, can cause suboptimal operation or even disruption to the V2G systems.
	\item \emph{Availability:} Data communication facilitates a number of functions in V2G systems. Therefore, its availability is crucial to provide seamless and efficient data transfer from mobile PEVs to fixed infrastructure. However, V2G communication can be disrupted, e.g., denial-of-service (DoS) attacks, which results in incomplete information to the V2G operator in operating the system.
\end{itemize}

\subsubsection{Solutions}

Given the above requirements, V2G communication infrastructure has to be designed and implemented accordingly. A few works have proposed different approaches to address different issues. The authors in~\cite{Yang2011P2} designed a security framework to protect the privacy of PEV users, thereby encouraging them to participate V2G systems. In the framework, all privacy information of PEV users and their aggregators are sent directly to a trusted authority. The trusted authority then adopts the ID-based restrictive partially blind signature technique to generate public/private key pairs, and sends them back to the PEV users and the aggregators. Based on these public/private key pairs, the aggregators can authenticate participated PEVs without knowing their identities while the PEV users can provide V2G services with secured information. As such, PEV users' information is protected from aggregators as well as from eavesdroppers since their information is encrypted by the trusted authority. As the method is relatively simple, its overhead is minimal. However, the system relies heavily on the trusted authority, which can become a single point of failure.

Different from~\cite{Yang2011P2}, the solutions proposed in~\cite{Liu2014Role} considered security for different states of vehicle's battery. In particular, the battery has three states, i.e., charging, fully-charged, and discharging. At each state, the PEV user has different security requirements such as identity, location, and energy status, and thus the corresponding security protocols were introduced. Similar to~\cite{Yang2011P2}, these protocols mainly focus on the authentication between PEV users and aggregators and the confidential information protection for PEV users. Nevertheless, in~\cite{Liu2014Role}, the authors also considered the data integrity issue for PEV users through using Hash functions together with signature algorithms. As such, the transmitted data from PEV users can be protected from malicious modification by cyber attackers. However, the solutions in~\cite{Liu2014Role} are more complicated and have considerable overheads.

In~\cite{Shuaib2016CognitiveRadio}, the authors discussed the jamming attack problem in smart grids as well as V2G systems. For such kind of networks, the useful information from service providers, e.g., energy price and locations of charging stations, may be unavailable to the PEV users due to diverse types of jamming attacks such as constant jamming, deceptive jamming, random jamming, and reactive jamming~\cite{Xu2005The feasibility}. The information unavailability problem can cause serious damage not only to the PEV users, but also to the service providers. On the one hand, the PEV users are unable to find the best charging station for charging/discharging to minimize the overall cost, e.g., traveling and energy costs. On the other hand, the service providers cannot maximize their profits because optimal economic policies cannot be applied to the PEVs, e.g., offering a low energy price in off-peak hours and/or for stations with redundant energy. Consequently, the PEV users may not be interested in participating in V2G systems due to the high cost, resulting in a significant revenue reduction to the V2G service providers. 

Different approaches were proposed in~\cite{Xu2004Channel} to deal with jamming attacks, namely channel surfing and spatial retreats. For the channel surfing approach, the wireless nodes will move their communications to another channel once jamming attacks are detected. For the spatial retreats, wireless nodes change their locations to outside the interference range of the jammers. Both approaches can mitigate the impact of the jamming attacks, but they are difficult to implement in V2G systems. This is from the fact that PEV users are mobile, and the communication channel between the PEV users and V2G systems are usually fixed. In~\cite{Hoang2015Performance}, a new solution based on the deception tactic to deal with smart jamming attacks was proposed. Basically, the core idea of the deception mechanism is using fake transmissions to undermine the attack ability of enemies, e.g., by wasting the energy of their adversaries. Thus, jammers may not be able to attack when V2G systems transmit actual information. Although this solution can effectively reduce adverse effects from smart jammers even when they use different attack strategies, it is inefficient if the jammers are powerful devices and have constant power supply. 

In practice, there are also many solutions proposed to address the jamming attacks in wireless networks as presented in~\cite{Pelechrinis2011Denial}. However, they can only reduce the impact of the attacks. A perfect solution which can completely avoid jamming attacks is impossible in practice. Hence, in this paper, we introduce a novel concept using cyber insurance to ``transfer'' cyber risks, e.g., unavailable information, of PEV users to a third party, e.g., a cyber insurance company. Under the insurance coverage, even without information about V2G systems, PEV users are always guaranteed the best price for charging/discharging. As a result, the PEV users' profits will be maximized, and thus they are encouraged to participate in V2G systems, yielding to a considerable revenue for the V2G service providers.

\section{Overview of Cyber Insurance}
\label{sec:O_CI}

In this section, we present an overview of cyber insurance. Cyber insurance is considered to be a promising solution to ``transfer'' risks from stackholders, i.e., the insured, to a third party, i.e., an insurer. Such risks include system failure and cyber attacks which can cause damage to PEV users.

\subsection{Definition, Fundamental Concepts, and Coverage}

With the prevalent applications of Internet-of-Things, everything can be connected to the Internet by wireline or wirelessly including V2G systems and PEVs. Internet has brought numerous advantages, but it also involves cyber risks including reliability and security. When such a connection is unavailable due to system failure or cyber attacks, not only financial losses, but also catastrophic danger to humans can happen. Hence, we need efficient and effective solutions to deal with cyber risks. Although there are many proposed reliable designs and security solutions, it was pointed out in~\cite{Pal2014Will} that it is impossible to achieve a perfect or near-perfect system reliability and cyber security protection. Therefore, cyber insurance can be considered to be a potential and efficient solution for cyber risk elimination and Internet security improvement.

\subsubsection{Definitions and fundamental concepts}

Cyber insurance can be defined in different contexts. For example, in the Internet context, cyber insurance is considered to be a set of policies that provide coverage against losses from Internet-related breaches in information security~\cite{Gordon2003AFrame}. In the business context, cyber insurance is a risk management technique via which network users' risks are transferred to an insurance company, in return for a fee~\cite{Pal2014Will}. In the market context, cyber insurance can be interpreted as a powerful tool to align market incentives towards improving Internet security~\cite{Majuca2005_The}. Therefore, in general, cyber insurance can be regarded as \emph{an insurance product that is used to protect businesses and individuals from cyber risks}. 

The followings are important fundamental concepts of cyber insurance.
\begin{itemize}
	\item \emph{Cyber risks:} are potential threats in the cyber world which can cause losses/damage to humans and society.
	\item \emph{Cyber insured:} is the user/customer who wants to be protected from cyber risks. 
	\item \emph{Cyber insurer:} is the insurance company which wants to take users' cyber risks together with a commensurate profit.
	\item \emph{Cyber insurance premium:} is the amount of money that the cyber insured has to pay to the cyber insurer to be protected.
	\item \emph{Cyber insurance contract:} is the signed deal between the cyber insured and the cyber insurer. 
	\item \emph{Claim:} is a formal request activated by the insured when a cyber risk has occurred. 
	\item \emph{Indemnity:} is the compensation from the cyber insurer to the insured for the loss/damage caused by cyber risks.
\end{itemize}

Basically, in a cyber insurance contract, the insured will agree to pay the insurance premium to the insurer in order to receive the protection from the insurer. In other words, the insured's risks are now ``transferred'' to the insurer, and the insurer can profit from the premium and efficient management of taking the risks. 

\subsubsection{Coverage}

Currently, cyber insurance covers losses and damage caused by cyber attacks to IT systems and the Internet. In general, cyber risks are categorized into two types, i.e., first-party and third-party, and thus cyber insurance policies are designed to cover either or both types of risks. In particular, the first-party insurance (www.abi.org.uk) covers the insured' own assets, and it involves: 
\begin{itemize}
	\item Losses or damage to digital assets
	\item Business interruption
	\item Cyber extortion
	\item Reputational damage 
	\item Theft of money or digital assets
\end{itemize}
Meanwhile, the third-party insurance covers the assets of subjects which are damaged by the insured. The third-party may involve:
\begin{itemize}
	\item Security and privacy breaches
	\item Multi-media liability
	\item Loss of third party data
	\item Third-party contractual indemnification
\end{itemize}

\subsection{Benefits of Cyber Insurance}

Cyber insurance has been considered to be an alternative solution to traditional security methods. In the following, we highlight and discuss benefits of cyber insurance in practice.  

\subsubsection{Benefits to the the insured} 

\begin{itemize}
	\item \emph{Mitigate damage:} By transferring risks to the insurer, the insured' damage will be significantly reduced when the risks happen. 
	\item \emph{Protected from insurers:} To avoid paying high compensation, the insurers have to make more efforts in implementing countermeasures to protect the insured.
	\item \emph{Improve self-defense:} The insured will be stimulated to implement self-protection methods in order to reduce the premium.
\end{itemize}

\subsubsection{Benefits to insurers} 

According to a recent report from the PwC Global State of Information Security Survey 2016, it was predicted that cyber insurance market will grow from \$2.5 billion in 2015 to \$7.5 billion by 2020~\cite{online_forbe}. This reveals that cyber insurance is a promising and attractive market because it will open many new business opportunities for insurers.

\subsubsection{Benefits to third-party and society} 

Unlike conventional insurance, cyber insurance requires insurers to have specialized knowledge about cyber security as well as network systems. This opens new opportunities for network security providers for consultation, support, and monitoring for insurers. Consequently, the development of cyber insurance results in a higher overall social welfare~\cite{Kesan2004The}.

\subsection{Implementation and Effectiveness of Cyber Insurance}

With aforementioned benefits, many applications of cyber insurance were implemented in practice especially for Internet security. In particular, in 1990, the first known cyber insurance policy was introduced by security software companies partnering with insurance companies in order to offer insurance-bundled software security services (software + insurance services)~\cite{Lelarge2009Economic}. The aim of these services is to not only mitigate losses, but also reduce residual risks for the insured. In 1998, the International Computer Security Association (ICSA Inc.) corporation introduced the hacker-related insurance packages, namely TRSecure service, to against hacker attacks to its clients~\cite{Poletti1998_First}. This is also known as the first stand-alone cyber insurance service which creates precedent for the development of later cyber insurance services of Lloyd's of London (https://www.lloyds.com/), AT\&T (http://www.mmc.com/), and AIG (www.aig.com)~\cite{Majuca2005_The}. 

Recently, the rapid growth of cyber insurance has been receiving a lot of attentions from the literature. Many research works have demonstrated the effectiveness as well as applicability of cyber insurance. In particular, in~\cite{Srinidhi2015Allocation}, the authors developed an analytical model for allocating optimal investments, and evaluated the role of cyber insurance in mitigating the influence on breach costs. Through analysis on impacts of insurance coverage, the authors showed that insurance is able to reduce over-investments for specific security-enhancing assets. Different from~\cite{Srinidhi2015Allocation}, the authors in~\cite{Ishikawa2016AStudy} adopted Monte Carlo simulation to evaluate the effectiveness of cyber insurance. In particular, the authors simulated a virtual company running the e-commerce site under cyber attacks and performed around 100 million simulation trials to estimate losses and evaluate efficiency of using cyber insurance. Through simulation results, the authors showed that cyber insurance can reduce the cost for the company up to 65\%. In addition, there are also some other research works studying the applicability as well as efficiency of using cyber insurance for software security~\cite{Laszka2015Should}, university networks~\cite{Saini2011Utility}, and Nigeria market~\cite{Adeleke2011Cyber}.

\subsection{Cyber Insurance Process}

\begin{figure}[h]
	\begin{center}
		$\begin{array}{c} \epsfxsize=3.4 in \epsffile{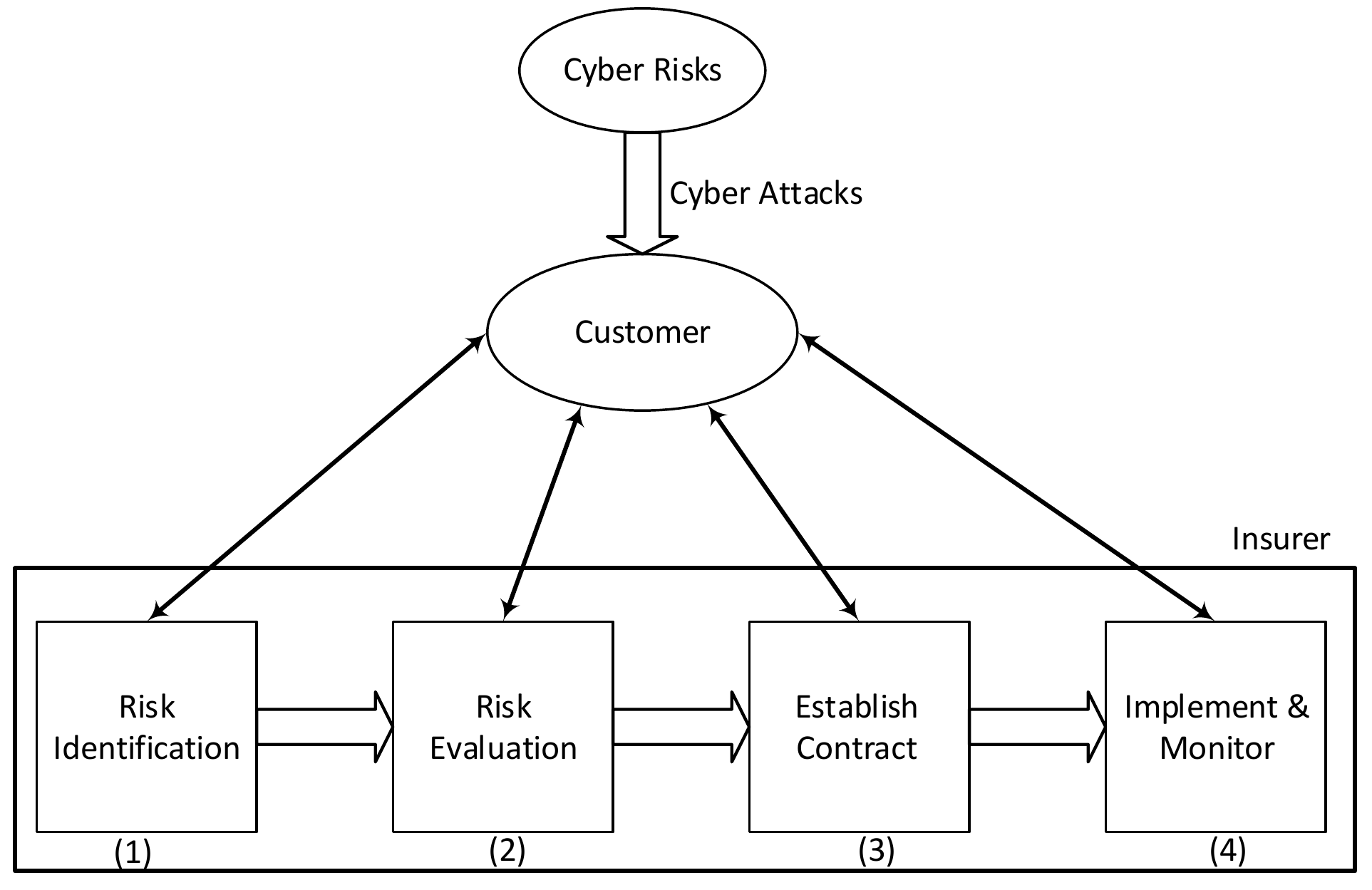} \\ [-0.2cm]
		\end{array}$
		\caption{Cyber insurance process.} 
		\label{fig:CI_process}
	\end{center}
\end{figure}

A cyber insurance process involves four main steps as illustrated in Fig.~\ref{fig:CI_process}. In the following, we will discuss step-by-step process of a cyber insurance. 

\subsubsection{Risk identification} 

This is the first step of a cyber insurance process.  After receiving a request from a customer, the insurer has to identify potential risks which may have negative impacts to the customer. To do so, the insurer needs to study the customer's coverage requirement, e.g., first-party and/or third-party coverage, then carries out investigations based on information provided by the customer to find threats and vulnerability of protected objects.

\subsubsection{Risk evaluation} 

In this step, the insurer will analyze and evaluate the risks by assessing the possibility of risks occurring as well as their potential damage. This is the most important step in the cyber insurance process because it will decide how to make a proper contract. If the insurer underestimates the risks, they will loose profits. By contrast, if the insurer overestimates the risks, the customer may not be interested in the insurance. However, in practice, this step is always the most difficult step in the cyber insurance process because it is often hard to estimate accurately the risks due to many reasons, e.g., asymmetry information between the insured and the insurer.

\subsubsection{Establish contract} 

After the risks are well investigated, the insurer proposes an insurance policy which prescribes terms, conditions, and exclusions for the insured. If the customer accepts this policy, a legal contract is signed between the insurer and the insured, i.e., the customer. On the other hand, if the customer disagrees with that offer, the insurer and the customer can negotiate to find a joint agreement. In the case if the customer does not accept any offers from the insurer, the process ends here.

\subsubsection{Implement and monitor}

Once the contract is made, the insurer will carry out solutions to protect the insured as well as to minimize its damage if cyber attacks happen. The solutions can include periodic monitoring and inspecting processes so as to make timely appropriate countermeasures if the risks occur. If the risks occur and cause losses to the insured, the insurer will verify the risks and handle claims from the insured as agreed in the contract.

\begin{table*}[!]
	\caption{Categories of cyber risks} 
	\label{table_taxo}
	\begin{centering}
		\begin{tabular}{|>{\centering\arraybackslash}m{1.2cm}|>{\centering\arraybackslash}m{1.7cm}|>{\centering\arraybackslash}m{6.2cm}|>{\centering\arraybackslash}m{5.8cm}|}
			\hline 
			\textbf{Category} & \textbf{Subcategory} & \textbf{Description} & \textbf{Elements} \tabularnewline
			\hline 
			\hline 
			\multirow{5}{*}{\parbox{1.2cm}{Actions of people}}
			& Inadvertent & Unintentional actions taken without malicious or harmful intent & Mistakes, errors, omissions 	\tabularnewline 	\cline{2-4} 
			& Deliberate & Actions taken intentionally and with intent to do harm & Fraud, sabotage, theft, and vandalism 	\tabularnewline	\cline{2-4} 
			& Inaction & Lack of action or failure to act in a given situation & Lack of appropriate skills, knowledge, guidance, and availability of personnel to take action  	\tabularnewline 	\cline{2-4} 
			\hline 
			\multirow{3}{*}{\parbox{1.2cm}{Systems and technology failures}}	
			& Hardware & Risks traceable to failures in physical equipment & Failure due to capacity, performance, maintenance, and obsolescence 	\tabularnewline 	\cline{2-4} 
			& Software & Risks stemming from software assets of all types, including programs, applications, and operating systems & Compatibility, configuration management, change control, security settings, coding practices, and testing 	\tabularnewline	\cline{2-4} 
			& Systems & Failures of integrated systems to perform as expected & Design, specifications, integration, and complexity 	\tabularnewline 	\cline{2-4} 
			\hline 
			\multirow{3}{*}{\parbox{1.2cm}{Failed internal processes}}	
			& Process design and/or execution & Failures of processes to achieve their desired outcomes due to poor process design or execution & Process flow, process documentation, roles and responsibilities, notifications and alerts, information flow, escalation of issues, service level agreements, and task hand-off 	\tabularnewline 	\cline{2-4} 
			& Process controls & Inadequate controls on the operation of the process & Status monitoring, metrics, periodic review, and process ownership 	\tabularnewline	\cline{2-4} 
			& Supporting processes & Failure of organizational supporting processes to deliver the appropriate resources & Staffing, accounting, training and development, and procurement  	\tabularnewline 	\cline{2-4} 
			\hline 
			\multirow{5}{*}{\parbox{1.2cm}{External events}}	
			& Catastrophes & Events, both natural and of human origin, over which the organization has no control and that can occur without notice & Weather event, fire, flood, earthquake, unrest 	\tabularnewline 	\cline{2-4} 
			& Legal issues & Risk arising from legal issues & Regulatory compliance, legislation, and litigation 	\tabularnewline 	\cline{2-4} 		
			& Business issues & Risks arising from changes in the business environment of the organization & Supplier failure, market conditions, and economic conditions 	\tabularnewline	\cline{2-4} 
			& Service dependencies & Risks arising from the organization's dependence on external parties & Utilities, emergency services, fuel, and transportation  	\tabularnewline 	\cline{2-4} 
			\hline 					
		\end{tabular}
		\par\end{centering}
\end{table*}

\subsection{Challenges and Solutions}

Although there are many benefits and applications, cyber insurance has to face some challenges which hinder its development. In the following, we discuss some important challenges and potential solutions proposed in the literature.

\subsubsection{Risk classification}

In the first step of the cyber insurance process, the insurer needs to identify the cyber risks which may cause losses to the customer and itself. However, different from the traditional insurance, cyber risks are diverse and there is currently no standard to classify and determine the cyber risks. In~\cite{Cebula2010ATaxonomy}, the authors presented the first taxonomy of operational cyber security risks with the aim to identify and organize the sources of operational cyber security risks. The taxonomy organizes the definition of operational risks into four main categories with elements and descriptions as shown in Table~\ref{table_taxo}. Although the empirical information about cyber risks in~\cite{Cebula2010ATaxonomy} is still relatively	 limited, the taxonomy provides the fundamental classification of cyber risks which is especially important in evaluating cyber risks in the second step of the cyber insurance process. 

\subsubsection{Risk assessment}

\begin{figure}[h]
	\begin{center}
		$\begin{array}{c} \epsfxsize=3.5 in \epsffile{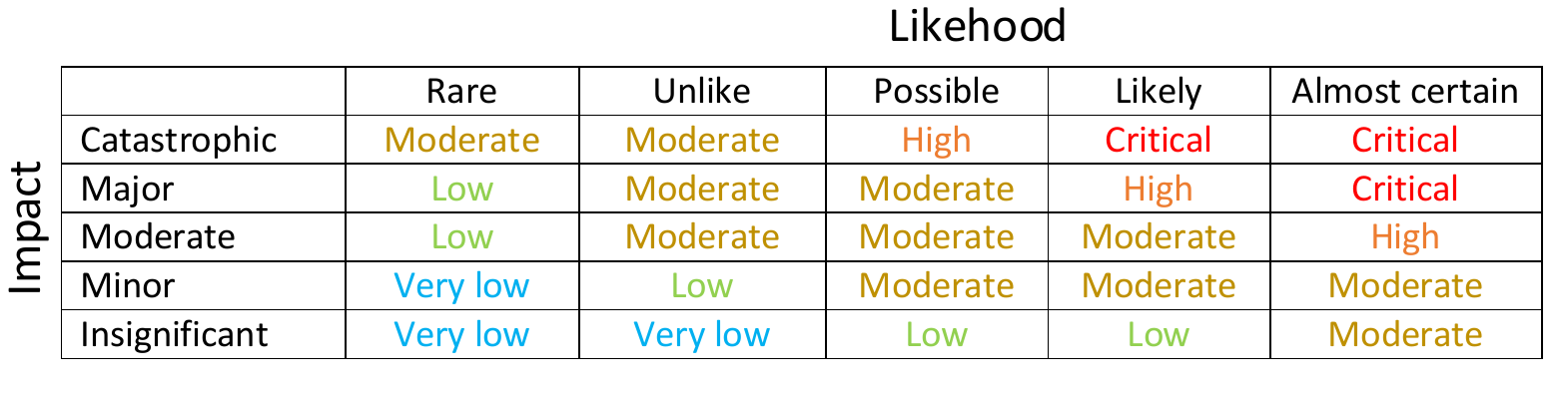} \\ [-0.2cm]
		\end{array}$
		\caption{Risk assessment matrix.} 
		\label{fig:RAM}
	\end{center}
\end{figure}

In the second step of the cyber insurance process, based on the risk analysis in the first step, the insurer needs to evaluate the risks in order to figure out an appropriate cyber insurance policy for the customer. To do so, one of the most common methods used in the literature as well as in practice is using Risk Assessment Matrix (RAM). The insurer can create a RAM to visualize the important areas of focus within their risk assessments, e.g., frequency, probability, severity, speed of development, and reputational impact as shown in Fig.~\ref{fig:RAM}. All of these factors serve as important guides in understanding the holistic nature of potential vulnerabilities and the probability of individual risks which impact the insured.

\subsubsection{Interdependent risks}

Another problem in evaluating cyber risks is the interdependence or correlated nature of the cyber-risks. Different from conventional insurance models, cyber insurance has to face the network security externalities due to the interdependence of entities. Specifically, cyber security of an entity depends on the operations as well as security levels of other entities in the network. To deal with this problem, insurance companies often impose insurance policies which do not cover such kind of risks. For example, in 2005, AIG offered cyber policies which exclude electric and telecommunication failures. However, this solution fails to prevent the infection spread, e.g., worms and virus, in the computer networks. In~\cite{Pal2010Analyzing}, the authors adopted a general mathematical framework to analyze policies of cooperative and non-cooperative Internet users under cyber-insurance coverage. An important conclusion drawn is that full insurance contracts encourage cooperative users to invest more for their self-defense, while partial insurance contracts motivate non-cooperative users to pay more for their self-defense mechanisms.

\subsubsection{Adverse selection}

In order to make a cyber insurance contract with the customer, the insurer must establish cyber insurance policies taking the adverse selection into consideration. In particular, adverse selection is an information asymmetry problem between the insured and the insurer where the insured has a complete awareness about his/her situation, while the insurer does not know, and thus it leads to the adverse selection problem for the insurer. To protect the insurer from this problem, insurance companies typically require their clients to have a current situation certification, e.g., life insurance companies require their clients to take certificated medical examinations. However, this problem becomes more difficult for cyber-risk insurance because there is currently no safety standardization for cyber systems. 

In order to deal with this problem, an insurance firm, called J.S.~Wurzler, proposed insurance contracts to cover damage caused by hackers' attacks with additional fee for clients using Microsoft's NT software~\cite{Gordon2003AFrame}. However, this is not an effective solution since cyber risks are not only governed by the insured's security system, but also by many cyber incidents, e.g., insured objects and their relations. As an effort to address this problem, the authors in~\cite{Elnagdy2016Cyber} introduced a model to link cyber incidents and risks with security insurance policies. Specifically, they developed a model, namely semantic cyber incident classification, which adopts semantic techniques to build a consistent and convincing knowledge representation for entities in cyber insurance system. Nevertheless, the authors did not consider all entities, and thus relations in cyber insurance need to be further investigated.

\subsubsection{Moral hazard}

The second major challenge in designing cyber insurance policies is moral hazard that refers to the problem when the insured under the insurance coverage relies on insurance contracts and pays less attention in preventing cyber risks. To prevent the insured from free-riding, a typical way is to issue additional terms for insurance contracts. For example, INSUREtrust (http://www.insuretrust.com/) offers a policy ``You agree to protect and maintain your computer system and your e-business information assets and e-business communications to the level or standard at which they existed and were presented...'', or Lloyd's of London insurance company requires ``The inured company maintains system security levels that are equal to or superior to those in place as at the inception of this policy''. 

However, these solutions do not encourage users in improving network security, thereby raising cyber risks for both the insurer and the insured. Thus, promotion policies can be used to handle this problem. For example, AIG provides discounts for clients who use Invicta Network's security devices or Lloyd's of London offers promotions for firms using Tripwire's Integrity security software. Nevertheless, different clients have different risk levels, and thus we cannot apply the same promotion for all clients. It was pointed out in~\cite{Pal2014Will} that for monopolistic cyber insurance contracts without client discrimination, there always exists an inefficient market in which the social welfare of users is not maximized at Nash equilibrium. However, if clients' discriminating premium policies are applied, the moral hazard problem is mitigated, thereby maximizing the overall network security.

\subsubsection{Setting premium}

This is the last step before an insurance contract is signed. There are two typical ways to determine the premium for an insurance contract in practice, i.e., through actuarial data and normative standards. However, both ways are unable to apply to the cyber insurance because cyber insurance is relatively new and there is currently no standard to establish cyber insurance premiums, while cyber actuarial data is not available since many companies are either unaware of a cyber attack or unwilling to disclose such attacks. Furthermore, there are also some other challenges in setting premiums for cyber contracts as pointed out in~\cite{Toregas2014Insurance}, e.g., underwriting process and premium-setting produces, and thus the authors suggested a research agenda developed by three main directions, i.e., policy, management, and technology.

\subsubsection{Other problems}

There are also other problems which have been also studied in the literature for the development of cyber insurance. For example, in~\cite{Pandey2014Applicability}, the authors examined the applicability of prediction markets~\cite{Wolfers2006Prediction} in forecasting and assessing information security events. In practice, prediction markets can be used as an efficient tool to improve aggregation of information, thereby improving the process of risk assessment and risk mitigation. In~\cite{Pandey2015Anovel}, a financial mechanism was introduced to incentivize coordinated efforts by security stakeholders in improving the information security ecosystem. The proposed solution is expected to address the problem of information asymmetry, negative externality and free riding for the insurer, and to negotiate a lower premium for the insured. In~\cite{Pal2013On}, a consumer pricing mechanism was examined to improve the profit for the insurer when a security vendor becomes a cyber-insurer. Through the simulation results, the authors showed that by using the proposed method, the security vendor's profit can be raised up to 25\%.

\begin{table*}[t]
	\caption{The expected payoff matrix} 
	\centering 
	\begin{tabular}{|c|c|c|}
		\hline
		&\textbf{Agent 2: Self-protection (S)}&\textbf{Agent 2: No-Protection (N)}\\
		\hline
		\hline
		\textbf{Agent 1: Self-protection (S)} & $u[w_0-c]$ & $(1-pq) u[w_0-c] + pq u[w_0-c-l]$ \\
		\hline
		\textbf{Agent 1: No-protection (N)} & $(1-p) u[w_0] + p u[w_0-l]$ & $p u[w_0-l] + (1-p) (pq u[w_0-l] + (1-pq) u[w_0]) $ \\
		\hline
	\end{tabular}
	\label{tab:expected_payoff}
\end{table*}

\subsection{Cyber Insurance Models}

Cyber risks are becoming more and more exacerbated to business and society, while countermeasures are still limited due to many reasons, e.g., information asymmetry and the complexity of cyber networks. Therefore, to attain efficient solutions, cyber insurance models which can quantify risks and measure effectiveness of cyber security and risk management strategies need to be taken into consideration. In this section, we discuss cyber insurance models with the aim of investigating the different characteristics offered by the insurer which tend to maximize the total outcome of the insurer as well as the insured.

\subsubsection{Classical model}

\begin{figure}[h]
	\begin{center}
		$\begin{array}{c} \epsfxsize=2.3 in \epsffile{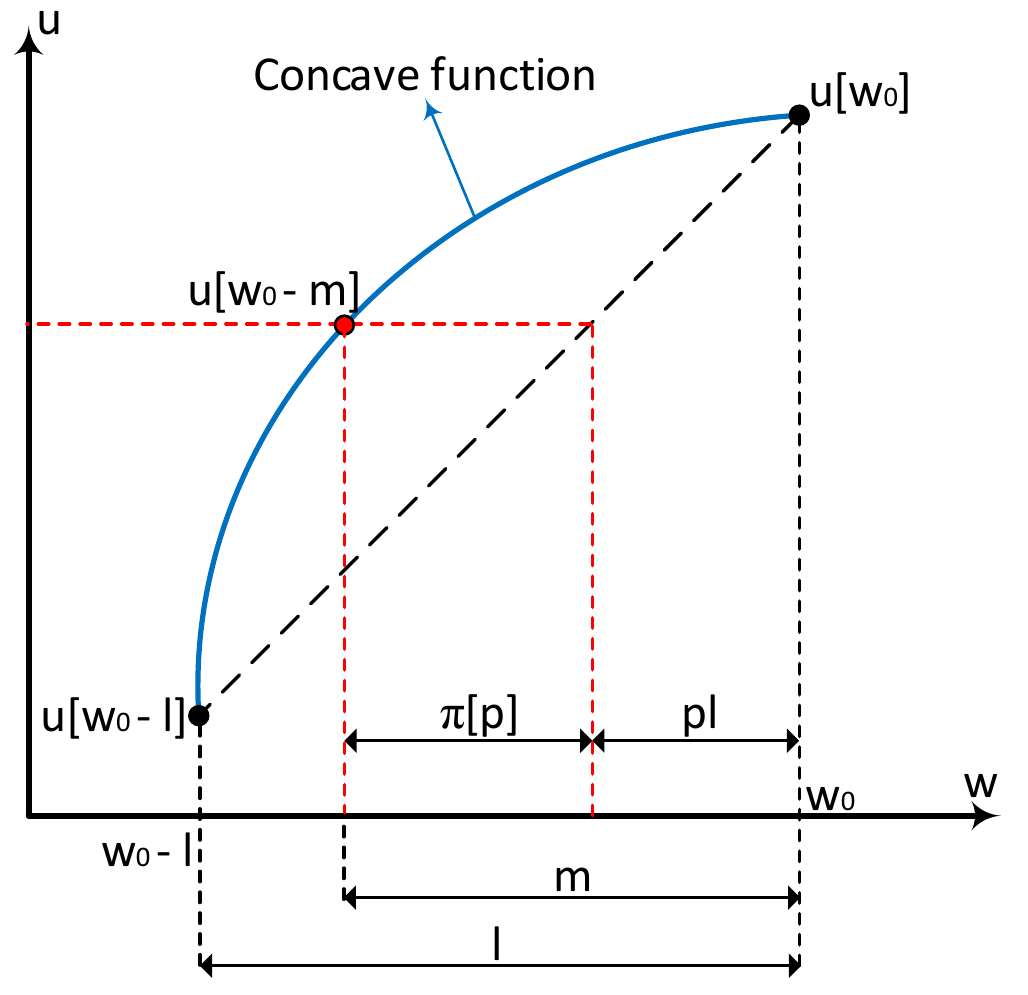} \\ [-0.2cm]
		\end{array}$
		\caption{Utility function.} 
		\label{fig:Classical_CI}
	\end{center}
\end{figure}

We consider a classical model for cyber insurance in which an agent (i.e., the insured) attempts to maximize its utility function $u[.]$. The agent is assumed to be rational and risk averse, i.e., its utility function is concave as shown in Proposition 2.1 in~\cite{Gollier2004Book}. We denote $w_0$ as the initial wealth of the agent, $\pi$ as the risk premium which is defined by the maximum amount of money that the agent is ready to pay to eliminate a pure risk $X$ (i.e., $\mathbb{E} (X) =0)$, $l$ as the potential loss of the agent caused by risk $X$ which is assumed to be a fixed value, and $p$ as the probability of loss. Then, the amount of money $m$ which the agent is ready to invest to eliminate the risk $X$ is derived as follows:
\begin{equation}
\label{eq:classical_premium_0}
p u[w_0 - l] + (1-p) u[w_0] = u[w_0-m]	.
\end{equation}
Then, from the results obtained in~\cite{Mossin1968Aspects}, we can derive the value of $m$ as follows:
\begin{equation}
\label{eq:classical_premium}
m = pl + \pi[p]		,
\end{equation}
where $\pi[p]$ is the risk premium when the loss probability equals $p$, and the term $pl$ represents the fair premium, i.e., the expected loss. The relation of terms in~(\ref{eq:classical_premium_0}) and in~(\ref{eq:classical_premium}) can be seen clearer in Fig.~\ref{fig:Classical_CI}.

For the classical cyber insurance model, $m$ can be expressed as the maximum acceptable premium for full coverage. This implies that if the insurer offers a full coverage with premium $\Omega$, the agent will accept the offer if $\Omega \leq m$. Thus, it can be observed that the premium $\Omega$ depends on the distribution of the loss, i.e., $p$ and $l$, and the existence of the insurance market will be determined by three parameters, i.e., $u$, $l$, and $p$. 

\subsubsection{Cyber insurance with self-protection}
\label{subsubsection: Cyber insurance with self-protection}

In~\cite{Bolot2008Cyber}, a cyber insurance model with self-protection for the insured was introduced. Different from the classical model where the agent has only two options, i.e., either purchase or do not purchase insurance, in the self-protection model, the agent has three options, i.e., self-protection, purchase insurance, or do not purchase. First, in the case without insurance, the agent has to decide whether to buy insurance or not. If we denote $c$ as the cost of self-protection and $p[c]$ as the corresponding probability of loss, we need to find the optimal value of $c^*$ to maximize the following utility function:
\begin{equation}
\label{eq:self_pro_CI}
\max_{c} f(c) = p[c] u[w_0-l-c] + (1-p[c]) u[w_0-c)]	.
\end{equation}
Obviously, when the agent invests money to protect itself, it will expect a lower probability of loss, and thus it is reasonable to assume that $p[c]$ is a non-increasing function of $c$. As a result, the optimization problem in~(\ref{eq:self_pro_CI}) has a unique solution, i.e., either $0$ or $c_t$, as demonstrated in~\cite{Bolot2008Cyber}. The authors then showed that if the cost for self-protection is less than a predefined threshold $c^{\dagger}$, then the agent will invest $c_t$ for self-protection. Otherwise, it will not invest for self-protection. 

Now, given the cyber insurance, the agent will have more choices. In the first case when $c<c^{\dagger}$, i.e., the agent will invest $c_t$ for self-protection, if the cost to buy insurance $c(\Omega)$ is less than $c_t$, the agent will buy insurance instead of investing for self-protection. Otherwise, if $c(\Omega)> c_t$, the agent will invest for self-protection only. In the second case when $c\geq c^{\dagger}$, i.e., the agent will not invest for self-protection, the model becomes the classical model where the agent has to decide to buy insurance or not, and we can use analysis in the previous section to find the optimal strategy for the agent. 

In general cases of cyber insurance with self-protection, the agent can choose a hybrid solution for self-protection and purchasing insurance. Specifically, the agent can invest a portion of cost, i.e., $\gamma c$, for self-protection, and the rest of cost, i.e., $(1-\gamma) c$, for insurance based on its demands. For example, for companies with good security system, it may invest more money for self-protection, and less money for insurance. In this case, the optimal value of $\gamma$ will be determined by the cost function of self-protection and insurance as shown in~\cite{Bolot2008Cyber}. However, for cyber insurance models with partial self-protection, the insurer has to face the moral hazard problem because when the agent is covered by insurance, it may take fewer measures to prevent losses. In this case, the insurer should tie up the premium to the amount of self-protection to avoid moral hazard behaviors from the insured~\cite{Ehrlich1972Market}. 

Obviously, cyber insurance models with self-protections bring more flexible and appropriate insurance policies for the agent compared with the classical model. Nevertheless, it was also highlighted in~\cite{Bolot2008Cyber} that there are still many difficulties as well as challenges in developing self-protection strategies in cyber insurance because the level of self-protection of the agent is still representing a complex and time-intensive task. 

\subsubsection{Interdependent model}

In~\cite{Kunreuther2003Interdependet}, the authors introduced a cyber insurance model for interdependent security (IDS) for the case with only two agents, and these agents have to face interdependent risk problem in the same network. In the IDS model, agents have to decide whether or not to invest in self-protection given a risk of losses which depends on the state of the other agents in the network. There are two causes of losses for an agent. The loss can be caused by an agent itself, i.e., direct loss, with probability $p$, and this loss can be caused by the other agents in the network, i.e., indirect loss, with probability $q$. Then, the utility function for these two agents can be determined as shown in Table~\ref{tab:expected_payoff}. Here, it is assumed that two agents are symmetric and $p$ and $q$ are independent parameters.

Denote $c_1=pl+ \pi[p]$ and $c_2 = p(1-pq)l + \pi[p+(1-p)pq] - \pi[pq]$, then by using game theory, the authors in~\cite{Bolot2008Cyber} showed the following results:
\begin{itemize}
	\item If $c\leq c_2$: The Nash equilibrium of the game is (S,S), i.e., both agents will invest in self-protection.
	\item If $c_2<c\leq c_1$: Both equilibria, i.e., (S,S) and (N,N), are possible and thus there is no Nash equilibrium solution for this game. 
	\item If $c_1<c$: The Nash equilibrium of the game is (N,N), i.e., both agents will not invest in self-protection.
\end{itemize}

Then, the authors integrated aforementioned analysis results into the insurance model in which the agents can choose whether to invest in self-protection and/or in a full coverage insurance. In this case, each agent will have three actions, i.e., purchase insurance, invest in self-protection, or do nothing, and similar to the case without insurance, the expected payoff matrix can be built and game theory can be adopted to analyze the Nash equilibrium solution for this IDS game with insurance. This model then can be extended to the case with $N$ agents with different kinds of network topoloty~\cite{Bolot2008Cyber} and/or to the case with partial insurance coverage~\cite{Pal2010Analyzing}.

There were also some other cyber insurance models studied in the literature. For example, the authors in~\cite{Pal2012Cyber} introduced a cyber insurance model to deal with the information asymmetry problem; Aegis model was introduced in~\cite{Pal2011CyberAegis} to deal with the case when the agent cannot discriminate between types of losses and risks; and Copulas was proposed in~\cite{Herath2011Copula} to forecast the value of losses and allow a proper pricing of cyber insurance. Each model has its own advantages and can be used in specific circumstances depending on the agent's situation.

\subsection{Evolution of Cyber Insurance Market}

Over the last two decades, the cyber insurance market has experienced great development steps with huge revenues for insurance companies. However, the cyber insurance market is still under the expectations. The reason is that cyber insurance companies mainly focus on exploiting conventional security market, i.e., Internet security market, which is gradually saturated due to the fierce competition among insurers. Thus, exploring new markets will be a potential solution for the development of cyber insurance in the future.

Recently, the rapid development of social networks and cloud computing has opened a great opportunity for cyber insurance. In particular, in early 2011, INSUREtrust  implemented the social media insurance package which allows social media companies to tailor the cover they buy to the risks they face. This insurance policy covers many problems related to the social networks such as defamation including libel and slander, intellectual property rights infringement, and so on. In 2013, the first cloud insurance platform was introduced by Cloudinsure (http://www.cloudinsure.com) to specifically address emerging privacy and security risks within the cloud environment. In the literature, there were a couple of research works proposing the idea of using cyber insurance to cloud security. In particular, the authors in~\cite{chaisiri2015} proposed a framework for cloud customers to manage the allocation of cloud security services and cyber insurance. The main aim of this framework is to maximize the profits for customers using cloud services, while minimizing their risks through insurance policies and their costs incurred in the process of using cloud services. Alternatively, a framework was introduced in~\cite{Gai2016ANovel} to reduce the implement cost, while remaining the security level for cyber insurance contracts. The core idea of~\cite{Gai2016ANovel} is using big data techniques to improve cyber security levels without a need of increasing financial budget. 

It is clear that there are still many potential markets which insurers can benefit, and this is the motivation for us to introduce a novel framework using cyber insurance in V2G systems. In the next section, we will show that cyber insurance is an efficient solution to address the cyber risks and optimize the benefit for PEV users. In addition, V2G systems are potential markets for cyber insurance companies.

\section{Risk Migration through Cyber Insurance in PEV Charging and Discharging}
\label{sec:RM_CI_PEV}

\subsection{System Model}

\subsubsection{PEV charging/discharging and V2G systems} 

We consider a V2G system in which a PEV user obtains the information about the energy price and the location of the charging stations through a V2G communication infrastructure. Different charging stations may have different prices at different time due to various factors, e.g., supply of renewable energy, consumer demand, and market influence. Therefore, based on the information provided by the V2G communication infrastructure, the PEV user can find the charging station which yields the lowest cost for charging or the highest profit for discharging. The cost for charging includes traveling cost and charging fee, while the profit for discharging equals the revenue obtained from discharging minus the traveling cost. 

Time is divided into $P$ periods, e.g., morning, afternoon, evening, and night. Thus, with the information about the charging stations, the cost (per unit of energy) to replenish energy for the PEV user in period $p$ is denoted by $c^{c}_p$, and $c^{c}_p \geq 0, \forall p = 1,\ldots,P$. Similarly, we denote by $c^{d}_p$ the discharging cost in the period $p$. However, different from $c^{c}_p$, $c^{d}_p \leq 0, \forall p = 1,\ldots,P$ since it represents the revenue of the PEV user. In practice, the information about charging stations may not be available to the PEV user for many reasons such as network failure and/or cyber attacks. Thus, if the PEV user decides to charge or discharge in period $p$ without information about charging stations, the cost for charging, denoted by $C^{c}_p$, could be higher, i.e., $c^{c}_p \leq C^{c}_p$, and the cost for discharging, denoted by $C^{d}_p$, could also be higher, i.e., $c^{d}_p \leq C^{d}_p$. 

Furthermore, we denote $l_p$ as the probability when the V2G communication infrastructure is unavailable in period $p$. Moreover, the PEV user has a battery with fixed capacity, denoted by $B$, and hence the energy storage is divided into $B$ levels, i.e., $1,2,\ldots, B$. 

\subsubsection{Cyber insurance for PEV charging and discharging} 

For V2G systems, when the information about charging stations is unavailable, there will be a risk to the PEV user. In particular, the PEV user may receive a higher cost for charging and a lower revenue for discharging. Therefore, we introduce the idea of using cyber insurance to transfer the risk from the PEV user, i.e., an insured, to an insurer who provides the price-guaranteed service. The insurer can be a third party, e.g., an insurance company, or in a form of extra services offered by the company owning charging stations and aggregators. The PEV user can buy the insurance by paying a premium, denoted by $m$. The insurer will then issue an insurance which is valid for a period of time to reserve the best price for the PEV user. In particular, if the PEV is under the insurance coverage, and it wants to be charged or discharged, the PEV user will pay the cost of $c^{c}_p$ or $c^{d}_p$, respectively, no matter whether the information about the charging stations is available or not. However, if the PEV user is not covered by the insurance and the information infrastructure is not available, the PEV user will incur the cost of $C^{c}_p$ or $C^{d}_p$, if it wants to charge or discharge, respectively. Again $c^{c}_p \leq C^{c}_p$ and $c^{d}_p \leq C^{d}_p$ as discussed in the previous section. 

\begin{figure}[h]
	\begin{center}
		$\begin{array}{c} \epsfxsize=3.5 in \epsffile{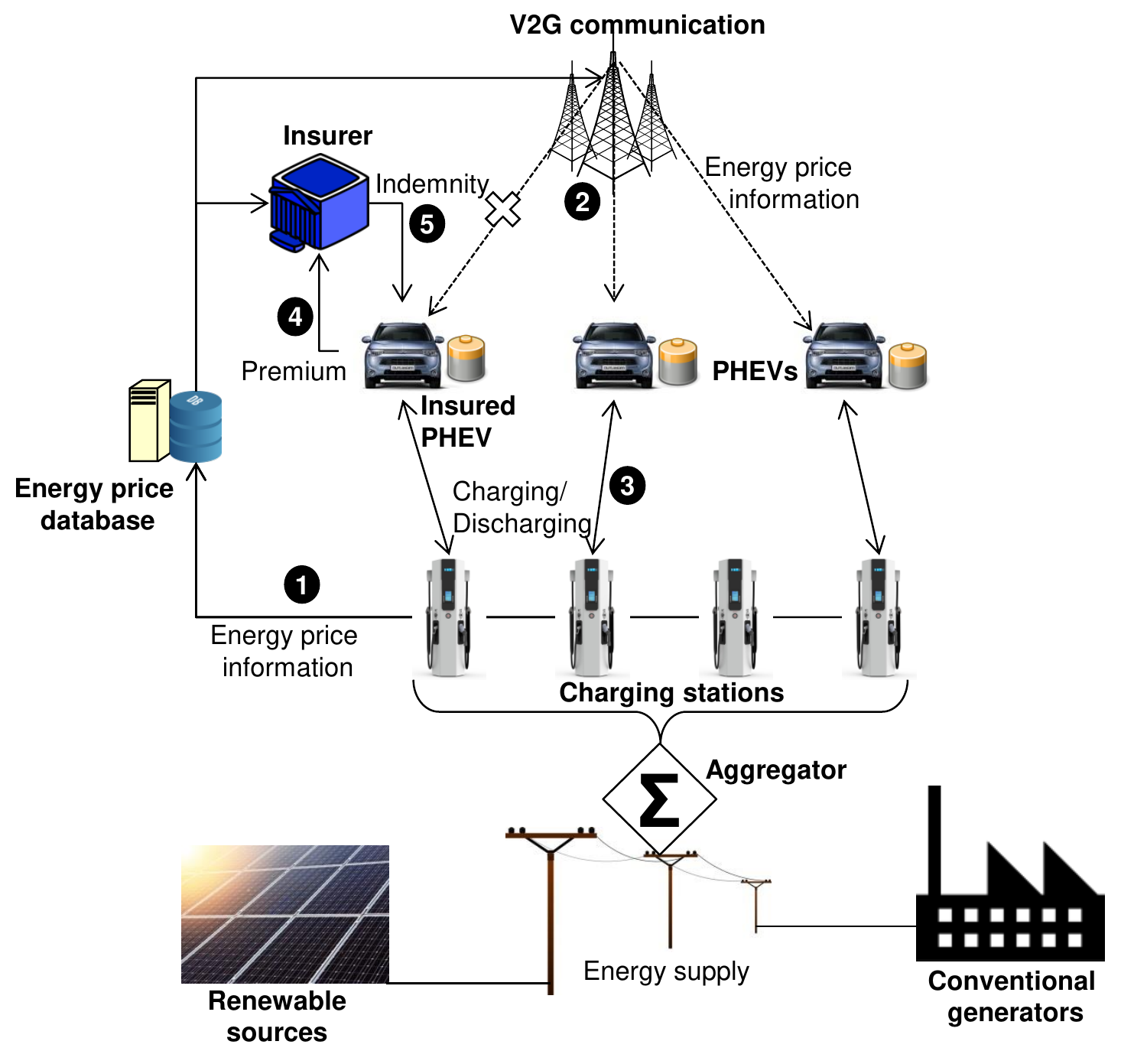} \\ [-0.2cm]
		\end{array}$
		\caption{Cyber insurance for PEV charging.} 
		\label{fig:System_Model}
	\end{center}
\end{figure}

In Fig.~\ref{fig:System_Model}, we show the system model of PEV charging/discharging and the cyber insurance which involves five main steps as follows.

\begin{itemize}
\item Firstly, the energy price information is collected from all charging stations at the energy price database.
\item Secondly, the information is transmitted to the PEV user through V2G communication channels.
\item Thirdly, the PEV user considers its battery level and uses the information to choose a suitable charging station. 
\item Fourthly, the PEV user can also choose to buy an insurance from the insurer by paying a certain premium to guarantee low charging fee and high discharging price.
\item Fifthly, if the V2G communication infrastructure is not available, the PEV user can still charge the battery with the guaranteed price while the extra cost is covered by the indemnity paid by the insurer.
\end{itemize}

From Fig.~\ref{fig:System_Model}, given the current state, the PEV user has to make two concurrent decisions. First, the PEV should charge, discharge, or do nothing in the current period. Second, the PEV should buy insurance or not. If the PEV buys insurance in this period, it will be guaranteed the best price for charging and discharging in next $\nu$ periods. The objective of the PEV user is to minimize the total cost, i.e., energy cost and insurance cost. To obtain optimal decisions, in the following, we will formulate a stochastic optimization problem based on Markov decision process (MDP).

\begin{figure*}[!]
	\begin{center}
		$\begin{array}{c} \epsfxsize=7.0 in \epsffile{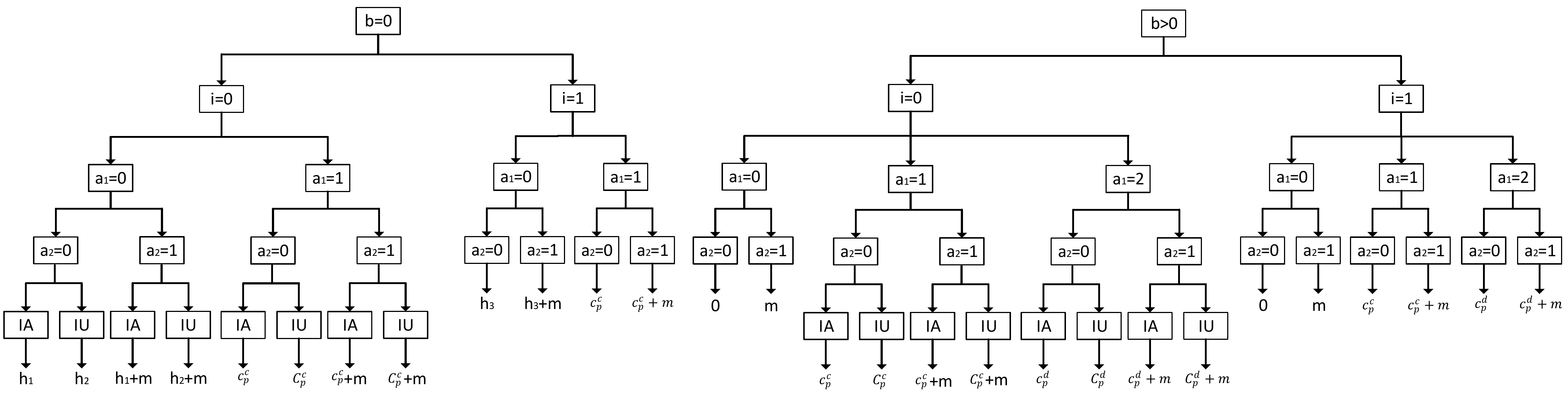} \\ [-0.2cm]
		\end{array}$
		\caption{Immediate cost function.} 
		\label{fig:cost_function}
	\end{center}
\end{figure*}

\subsection{Problem Formulation}

\subsubsection{State space} 

We define the state space of the PEV user as follows:
\begin{equation}
\mathcal{S} \triangleq   \mathcal{B} \times \mathcal{P} \times \mathcal{I} ,
\end{equation}
where $\times$ is the Cartesian product, $b \in \mathcal{B} = \{1,\ldots,B\}$ is the battery level of the PEV user, $p \in \mathcal{P} = \{1,\ldots,P\}$ represents the time period, and $i \in \mathcal{I} = \{0,1\}$ expresses the current insurance status of the PEV user. Thus, the state of the PEV user is then defined as a composite variable $s=(b,p,i) \in \mathcal{S}$. 

\subsubsection{Action space}

The action space is defined by:
\begin{equation}
\mathcal{A} \triangleq \mathcal{A}_1 \times \mathcal{A}_2	,
\end{equation}
where $a_1 \in \mathcal{A}_1 = \{0,1,2\}$, $a_2 \in \mathcal{A}_2= \{0,1\}$, and they can be defined as follows:
\begin{equation}
a_1 = \left\{
\begin{array}{ll}
0, & \text{if the PEV user does neither charging nor} \\
     &  \text{discharging},	\\
1, & \text{if the PEV user performs charging},	\\
2, & \text{if the PEV user performs discharging},
\end{array}
\right.
\end{equation}
and 
\begin{equation}
a_2 = \left\{
\begin{array}{ll}
0, & \text{if the PEV user does not buy insurance} 	,	\\
1, & \text{if the PEV user buys insurance} 	.
\end{array}
\right.
\end{equation}

While choosing $a_2$ depends on the demand of the PEV user only, i.e., the PEV can choose either to buy or not to buy at any period without concerning its current state, $a_1$ must be selected based on the current state of the PEV user. For example, when the current battery level is zero, the PEV user cannot choose action ``discharging''. Therefore, the action space $A_1$ can be redefined as follows: 
\begin{equation}
\label{eq:}
\mathcal{A}_{1} = \left\{
\begin{array}{ll}
\{0,1\}, & \text{if $b=0$, }\\
\{0,1,2\}, & \text{if $b>0$ and $b<B$, }\\
\{0,2\}, & \text{if $b=B$. }\\
\end{array}
\right.
\end{equation}

\subsubsection{Immediate cost function}

We denote $f_c$ as the immediate cost function for the PEV user, and it can be defined depending on different cases as shown in Fig.~\ref{fig:cost_function}. In Fig.~\ref{fig:cost_function}, when the battery level is zero, i.e., $b=0$, if the PEV user takes action ``do nothing'', i.e., $a_1=0$, then the PEV user will receive a heavy cost $h_1$ or $h_2$ corresponding to the cases when the PEV user is not under or under the insurance coverage, respectively. These costs are to prevent the PEV user from the energy depletion status, and in general we have $h_2>h_1$. In Fig.~\ref{fig:cost_function}, IA and IU stand for ``insurance is available'' and ``insurance is unavailable'', respectively. 

In this paper, we aim to find the optimal policy $\Psi^*$ to minimize the expected average cost of the PEV user in a long run which can be defined as follows:
\begin{equation}
\min_{\Psi} 	\mathcal{C} (\Psi) = \lim_{T \rightarrow \infty}  \frac{1}{T} \mathbb{E}_{\Psi} \Bigg[ \sum_{t=1}^{T} f_c \Big(s_t,\Psi(a_t)	\Big)	\Bigg],
\end{equation}
where $s_t$ and $a_t$ are the state and action at the $t$-th time period, respectively.

\subsection{Optimal Policy with Learning Algorithm}

In our considered system, cyber risks are random and unpredicted, and thus it is intractable to estimate the probability of cyber risks at each time period, i.e., $l_p$. As a result, we are unable to derive the transition probability matrix to find the optimal policy for the PEV user. Therefore, in this section, we introduce a learning algorithm based on the simulation-based method to help the PEV user make optimal decisions in an online fashion. 

\subsubsection{Parameterized policy}

We consider a randomized parameterized policy which is well studied in the literature~\cite{Marbach2001, Olivier2007, Baxter2001}. Under the randomized parameterized policy, when the PEV user is at state $s$, it will select action $a$ with the probability $\mu_{\Theta}(s,a)$ as follows:
\begin{equation}
\label{eq:randomized_parameterized_function}
\mu_{\Theta}(s,a) = \frac{\exp (\theta_{s,a})}{\sum_{a_i \in \mathcal{A}} \exp(\theta_{s,a_i})}	,
\end{equation}
where $\Theta = \{\theta_{s,a} \in \mathbb{R} \}$ is the parameter vector of the PEV user. Furthermore, every $\mu_{\Theta}(s,a)$ must not be negative and $\sum_{a\in \mathcal{A}} \mu_{\Theta} (s,a) =1$. 

Under the randomized parameterized policy $\mu_{\Theta}(s,a)$, the transition probability function will be parameterized as follows:
\begin{equation}
p_b(s'|s,\Psi(\Theta)) = \sum_{a \in \mathcal{A}} \mu_{\Theta}(s,a)	p_b(s'| s, a)	,
\end{equation}
for all $s, s' \in \mathcal{S}$, and $p_b(s'| s, a)$ is the transition probability from state $s$ to state $s'$ when action $a$ is taken. Similarly, we have the parameterized immediate cost function defined as follows:
\begin{equation}
\label{eq:average_throguhput_defination}
f_c (s, \Theta)	= \sum_{a \in \mathcal{A}} \mu_{\Theta}(s,a) f_c (s, a) .
\end{equation}

Our objective is to minimize the average cost of the PEV user under the randomized parameterized policy $\mu_{\Theta}(s,a)$, which is denoted by $\Psi(\Theta)$. Then we make some necessary assumptions as follows.
\begin{assumption}
\label{ass:recurrent_state}
The Markov chain is aperiodic and there exists a state $s^{*}$ which is recurrent for each of such Markov chain. 
\end{assumption}

\begin{assumption}
\label{ass:derivatives}
For every state pair $s, s' \in \mathcal{S}$, the transition probability function $p_b(s'|s,\Psi(\Theta))$ and the immediate cost function $f_c (s, \Theta)$ are bounded, twice differentiable, and have bounded first and second derivatives. 
\end{assumption}

Assumption~\ref{ass:recurrent_state} implies that the system has a Markov property, and Assumption~\ref{ass:derivatives} ensures that the transition probability function and the immediate cost function depend ``smoothly'' on the parameter vector $\Theta$. Then, we can define the parameterized average cost (i.e., the cost under the parameter vector $\Theta$) by
\begin{equation}
\mathcal{C} (\Theta) = \lim_{T\rightarrow \infty} \frac {1}{T} \mathbb{E}_{\Theta} \Big[ \sum_{t=0}^{T} f_c (s_t, \Theta)\Big] 	,
\end{equation}
where $s_t$ is the state of the PEV user at time step $t$. $\mathbb{E}_{\Theta}[ \cdot ]$ is the expectation under parameter vector $\Theta$. Under Assumption~\ref{ass:recurrent_state}, the average cost $\mathcal{C}(\Theta)$ is well defined for every $\Theta$, and does not depend on the initial state $\Theta_0$. Moreover, we have the following balance equations
\begin{eqnarray}
\label{eq: balance equation}
& & \sum_{s \in \mathcal{S}} \pi_{\Theta}({s}) p_b(s'|s,\Psi(\Theta)) = \pi_{\Theta}({s'}), \forall  s' \in \mathcal{S} 	,	\nonumber\\
& & \sum_{s \in \mathcal{S}} \pi_{\Theta}({s})  = 1	,
\end{eqnarray}
where $\pi_{\Theta}({s})$ is the steady-state probability of state $s$ under the parameter vector $\Theta$. These balance equations have a unique solution defined as a vector $\Pi_{\Theta} = \left[	\begin{array}{ccc}	\cdots	&	\pi_{\Theta}({s})	&	\cdots	\end{array}	\right]^\top$. Then, the average cost can be expressed as follows:
\begin{equation}
\label{eq:average throughput}
\mathcal{C} (\Theta) = \sum_{s \in \mathcal{S}} \pi_{\Theta}({s}) f_c (s, \Theta)	.
\end{equation}

\subsubsection{Policy gradient method}

We define the differential cost $d(s,\Theta)$ at state $s$ by
\begin{equation}
d(s, \Theta) = \mathbb{E}_{\Theta} \left[ \sum_{t=0}^{T^{\dagger}-1} \left( 	f_c (s_t, \Theta) - \mathcal{C} (\Theta)	\right) | s_{0} = s \right]	,
\end{equation} 
where $T^{\dagger}=\min\{t>0 | s_t = s^{*}\}$ is the first future time that state $s^{*}$ is visited. Here, it is worth to note that, the main aim of defining the differential cost $d(s, \Theta)$ is to represent the relation between the average cost and the immediate cost at state $s$, instead of the recurrent state $s^{*}$. Additionally, under Assumption~\ref{ass:recurrent_state}, the differential cost $d(s, \Theta) $ is a unique solution of the Bellman equation defined as follows:
\begin{equation}
d(s, \Theta)  = f_c (s, \Theta) - \mathcal{C} (\Theta)+\sum_{s' \in \mathcal{S}} p_b (s'|s,\Psi(\Theta)) d(s', \Theta)	,
\end{equation}
for all $s \in \mathcal{S}$. Then, we propose Proposition~\ref{prop:prop_policy_gradient} to calculate the gradient of the average cost as follows:
\begin{proposition}
	\label{prop:prop_policy_gradient}
	Let Assumption~\ref{ass:recurrent_state} and Assumption~\ref{ass:derivatives} hold, then \\
	\begin{equation}
	\nabla \mathcal{C} (\Theta) = \sum_{s \in \mathcal{S}} \pi_{\Theta}(s) \Big(\nabla f_c (s, \Theta) + \sum_{s' \in \mathcal{S}} \nabla p_b (s'|s,\Psi(\Theta)) d(s', \Theta)  \Big) 	.
	\end{equation}
\end{proposition}
Proposition~\ref{prop:prop_policy_gradient} represents the gradient of the average cost $C(\Theta)$, and the proof of Proposition~\ref{prop:prop_policy_gradient} is provided in Appendix~\ref{appendix:prop:prop_policy_gradient}.

\subsubsection{An idealized gradient algorithm}

Using Proposition~\ref{prop:prop_policy_gradient}, we can formulate the idealized gradient algorithm based on the form proposed in~\cite{Bertsekas_1995_Nonlinearprogramming} given as follows:
\begin{equation}
\label{idealized_algorithm_theta}
\Theta_{t+1} = \Theta_{t} - \rho_{t} \nabla \mathcal{C} (\Theta_{t})	,
\end{equation}
where $\rho_{t}$ is a step size and $\nabla \mathcal{C} (\Theta_{t})$ is the gradient of average cost function. Under a suitable step size satisfying Assumption~\ref{ass:step size} and Assumption~\ref{ass:recurrent_state} is hold, it is proved that $\lim_{t \rightarrow \infty} \nabla \mathcal{C}(\Theta_{t}) = 0$ and thus $\mathcal{C}(\Theta_{t})$ converges~\cite{Bertsekas_1995_Nonlinearprogramming}.

\begin{assumption}
	\label{ass:step size}
	The step size $\rho_{t}$ is deterministic, nonnegative and satisfies the following conditions,
	\begin{equation}
	\sum_{t=1}^{\infty}\rho_{t} = \infty, \mbox{ and } \sum_{t=1}^{\infty} ( \rho_{t} )^{2} < \infty	.
	\end{equation}
\end{assumption}

\subsubsection{Learning algorithm}

The idealized gradient method can minimize the average cost $\mathcal{C} (\Theta)$, if we can calculate the gradient of the function $\mathcal{C} (\Theta_{t})$ with respect to $\Theta$ at each time step. However, if the system has a large state space, it is impossible to compute the exact gradient of $\mathcal{C} (\Theta_{t})$. Therefore, we alternatively consider an approach that can estimate the gradient of $\mathcal{C} (\Theta_{t})$ and update parameter vector $\Theta$ accordingly in an online fashion. 

Since $\sum_{a\in \mathcal{A}} \mu_{\Theta} (s,a) =1$, we can derive that $\sum_{a\in \mathcal{A}} \nabla \mu_{\Theta} (s,a) = 0$ for every $\Theta$. From~(\ref{eq:average_throguhput_defination}), we have 
\begin{equation}
\begin{aligned}
\nabla f_c (s, \Theta)  = & \sum_{a \in \mathcal{A}} \nabla \mu_{\Theta}(s,a) f_c (s, a)	\\
= &	\sum_{a \in \mathcal{A}} \nabla \mu_{\Theta}(s,a) \big( f_c (s, a)  - \mathcal{C} (\Theta)  \big) 	,
\end{aligned}
\end{equation}
since $\sum_{a\in \mathcal{A}} \nabla \mu_{\Theta} (s,a) = 0$.

Moreover, we have 
\begin{equation}
\begin{aligned}
&\sum_{s' \in \mathcal{S}} p_b (s'|s,\Psi(\Theta)) d(s', \Theta) \\ 
& = \sum_{s' \in \mathcal{S}} \sum_{a \in \mathcal{A}} \nabla \mu_{\Theta} (s,a) p_b (s'|s,a)  d(s',\Theta)	,
\end{aligned}
\end{equation}
for all $s \in \mathcal{S}$.

Therefore, along with Proposition~\ref{prop:prop_policy_gradient}, we can derive the gradient of $\mathcal{C} (\Theta)$ as follows:
\begin{equation}
\begin{aligned}
\nabla \mathcal{C} (\Theta)  = &	\sum_{s \in \mathcal{S}} \pi_{\Theta}({s}) \Big(\nabla f_c (s, \Theta) + \sum_{s' \in \mathcal{S}} \nabla p_b (s'|s,\Psi(\Theta)) d(s', \Theta) \Big) \\
= &	\sum_{s \in \mathcal{S}} \pi_{\Theta}({s})  \Big( \sum_{a \in \mathcal{A}} \nabla \mu_{\Theta}(s,a) \big( f_c (s, a)  - \mathcal{C} (\Theta)  \big)	\nonumber	\\
&	+ \sum_{s' \in \mathcal{S}} \sum_{a \in \mathcal{A}} \nabla \mu_{\Theta} (s,a) p_b (s'|s,a)  d(s',\Theta)	 \Big) \\
= &	\sum_{s \in \mathcal{S}} \pi_{\Theta}({s}) \sum_{a \in \mathcal{A}} \nabla \mu_{\Theta}(s,a)  \Big(  \big( f_c (s, a)  - \mathcal{C} (\Theta)  \big) \nonumber	\\
&	+ \sum_{s' \in \mathcal{S}} p_b (s'|s,a)  d(s',\Theta)	 \Big) \\ 
= &	\sum_{s \in \mathcal{S}} \sum_{a \in \mathcal{A}} \pi_{\Theta}({s}) \nabla \mu_{\Theta}(s,a) q_{\Theta}(s,a),
\end{aligned}
\end{equation}
where 
\begin{equation}
\begin{aligned}
q_{\Theta}(s,a)&= \big( f_c (s, a)  - \mathcal{C} (\Theta)	\big) + \sum_{s' \in \mathcal{S}} p_b (s'|s,a)  d(s',\Theta)	 \\
&= \mathbb{E}_{\Theta} \Bigg[\sum_{t=0}^{T^{\dagger}-1}\big( f_c ( s_t, a_t) -\mathcal{C} (\Theta) \big) | s_{0} = s, a_{0}=a \Bigg].
\end{aligned}
\end{equation}
Here, $q_{\Theta}(s,a)$ can be interpreted as the differential cost if action $a$ is taken based on policy $\mu_{\Theta}$ at state $s$. Then, we present Algorithm~\ref{algorithm0} that updates the parameter vector $\Theta$ at the visits to the recurrent state $s^{*}$. 

\begin{algorithm}
	\caption{Algorithm to update the parameter vector $\Theta$ at the visits to the recurrent state $s^{*}$}
	\label{algorithm0}
	At the time step $t_{m+1}$ of the $(m+1)$th visit to state $s^{*}$, we update the parameter vector $\Theta$ and the estimated average cost $\widetilde{\psi}$ as follows: \\
	\begin{equation}
	\Theta_{m+1} = \Theta_{m} - \rho_{m} F_{m} (\Theta_{m},\widetilde{\psi}_{m}),
	\end{equation}
	\begin{equation}
	\widetilde{\psi}_{m+1} = \widetilde{\psi}_{m} + \kappa \rho_{m}\sum_{t'=t_{m}}^{t_{m+1}-1}\Big( f_c (s_{t'}, a_{t'}) - \widetilde{\psi}_{m} \Big)	,
	\end{equation}
	where
	\begin{equation}
	F_{m}(\Theta_{m},\widetilde{\psi}_{m}) = \sum_{t'=t_{m}}^{t_{m+1}-1} \widetilde{q}_{\Theta_{m}}(s_{t'},a_{t'}) \frac{\nabla \mu_{\Theta_{m}}(s_{t'},a_{t'})}{\mu_{\Theta_{m}}(s_{t'},a_{t'})},
	\end{equation}
	\begin{equation}
	\widetilde{q}_{\Theta_{m}}(s_{t'},a_{t'}) = \sum_{t=t'}^{t_{m+1}-1}\Big( f_c (s_{t}, a_{t}) - \widetilde{\psi}_{m}\Big).
	\end{equation}
\end{algorithm}

In \textbf{Algorithm~\ref{algorithm0}}, $\kappa$ is a positive constant and $\rho_{m}$ is a step size that satisfies Assumption~\ref{ass:step size}. The term $F_{m}(\Theta_{m},\widetilde{\psi}_{m})$ represents the estimated gradient of the average cost, and it is calculated by the cumulative sum of the total estimated gradient of the average cost between two successive visits (i.e., the $m$th and $(m+1)$th visits) to the recurrent state $s^*$. Furthermore, $\nabla \mu_{\Theta_{m}}(s_{t'},a_{t'})$ expresses the gradient of the randomized parameterized policy function that is provided in~(\ref{eq:randomized_parameterized_function}). \textbf{Algorithm~\ref{algorithm0}} enables us to update the parameter vector $\Theta$ and the estimated average cost $\widetilde{\psi}$ iteratively. Accordingly, we derive the following convergence result for \textbf{Algorithm~\ref{algorithm0}}.

\begin{proposition}
	\label{prop2}
	Let Assumption~\ref{ass:recurrent_state} and Assumption~\ref{ass:derivatives} hold, and let $(\Theta_{0}, \Theta_{1}, \ldots, \Theta_{\infty})$ be the sequence of the parameter vectors generated by Algorithm~\ref{algorithm0} with a suitable step size $\rho$ satisfying Assumption~\ref{ass:step size}, then  $\psi({\Theta_{m}})$ converges and 
	\begin{equation}
	\lim_{m\rightarrow \infty} \nabla \mathcal{C} (\Theta_{m}) = 0,
	\end{equation}
	with probability one. 
\end{proposition}
The proof of \textbf{Proposition~\ref{prop2}} is given in \textbf{Appendix~\ref{appendix:prop2}}.

\subsubsection{Online learning algorithm}

In \textbf{Algorithm~\ref{algorithm0}}, to update the value of the parameter vector $\Theta$ at the next visit to the state $s^{*}$, we need to store all values of $\widetilde{q}_{\Theta_{m}}(s_{t'},a_{t'})$ and $\frac{\nabla \mu_{\Theta_{m}}(s_{t'},a_{t'})}{\mu_{\Theta_{m}}(s_{t'},a_{t'})}$ between two successive visits. However, this method could result in a slow processing. Therefore, we modify \textbf{Algorithm~\ref{algorithm0}} to improve the efficiency. First, we rewrite $F_{m}(\Theta_{m},\widetilde{\psi}_{m})$ as follows:
\begin{equation}
\begin{aligned}
F_{m}(\Theta_{m},\widetilde{\psi}_{m}) & = \sum_{t'=t_{m}}^{t_{m+1}-1} \widetilde{q}_{\Theta_{m}}(s_{t'},a_{t'}) \frac{\nabla \mu_{\Theta_{m}}(s_{t'},a_{t'})}{\mu_{\Theta_{m}}(s_{t'},a_{t'})} , \\
&= \sum_{t'=t_{m}}^{t_{m+1}-1}   \frac{\nabla \mu_{\Theta_{m}}(s_{t'},a_{t'})}{\mu_{\Theta_{m}}(s_{t'},a_{t'})}  \sum_{t=t'}^{t_{m+1}-1} \big( f_c (s_{t}, a_{t}) - \widetilde{\psi}_{m}\big) , \\
&= \sum_{t=t_{m}}^{t_{m+1}-1} \big( f_c (s_{t}, a_{t}) - \widetilde{\psi}_{m}\big)  z_{t+1}, 
\end{aligned}
\end{equation}
where
\begin{equation}
z_{k+1} = \left\{ 
\begin{array}{ll}
\frac{\nabla \mu_{\Theta_{m}}(s_{t},a_{t})}{\mu_{\Theta_{m}}(s_{t},a_{t})}, & \text{if} \phantom{1} t = t_{m},\\
z_{t} + \frac{\nabla \mu_{\Theta_{m}}(s_{t},a_{t})}{\mu_{\Theta_{m}}(s_{t},a_{t})}, & t=t_{m}+1,\ldots,t_{m+1}-1. \\
\end{array}
\right.
\label{eq:zupdate_time}
\end{equation}

We then derive \textbf{Algorithm~\ref{algorithm1}}, which is able to update the parameter vector $\Theta$ at each time step as follows:

\begin{algorithm}
\caption{Algorithm to update $\Theta$ at each time step}
\label{algorithm1}
	At time step $t$, the state is $s_t$, and the values of $\Theta_t$, $z_t$, and $\widetilde{\psi}(\Theta_t)$ are available from the previous iteration. We update $z_t$, $\Theta_t$, and $\widetilde{\psi}$ according to: \\
	\begin{equation}
	z_{t+1} = \left\{ 
	\begin{array}{ll}
	\frac{\nabla \mu_{\Theta_{t}}(s_{t},a_{t})}{\mu_{\Theta_{t}}(s_{t},a_{t})}, & \text{if} \phantom{1} s_{t} = s^{*}\\
	z_{t}+\frac{\nabla \mu_{\Theta_{t}}(s_{t},a_{t})}{\mu_{\Theta_{k}}(s_{t},a_{t})}, & \text{otherwise,} \\
	\end{array}
	\right.
	\label{eq:zupdate_state}
	\end{equation}
	\begin{equation}
	\Theta_{t+1} = \Theta_{t} - \rho_{t} \big( f_c (s_t, a_t) -\widetilde{\psi}_{t} \big) z_{t+1}, 
	\end{equation}
	\begin{equation}
	\widetilde{\psi}_{t+1} = \widetilde{\psi}_{t} + \kappa\rho_{t} \big( f_c (s_t, a_t) - \widetilde{\psi}_{t} \big).
	\end{equation}
\end{algorithm}

In \textbf{Algorithm~\ref{algorithm1}}, $\kappa$ is a positive constant, $\rho_{t}$ is the step size of the algorithm, and $\widetilde{\psi}_{t}$ can be expressed as the estimated average cost of the PEV user at time step $t$. 

\subsection{Performance Evaluation}
\label{sec: Performance Evaluation}

In this section, we perform simulations using MATLAB to evaluate the performance of the proposed solution. We first show the impact of the infrastructure information unavailability to the cost of the PEV user. We then evaluate the benefits of using cyber insurance for the V2G system. We will show that, by using cyber insurance, the PEV user can reduce her average cost for charging and increase her average profit for discharging as well. 

\subsubsection{Cost due to V2G communication infrastructure unavailability}

\begin{figure}[h]
	\begin{center}
		$\begin{array}{c} \epsfxsize=2.6 in \epsffile{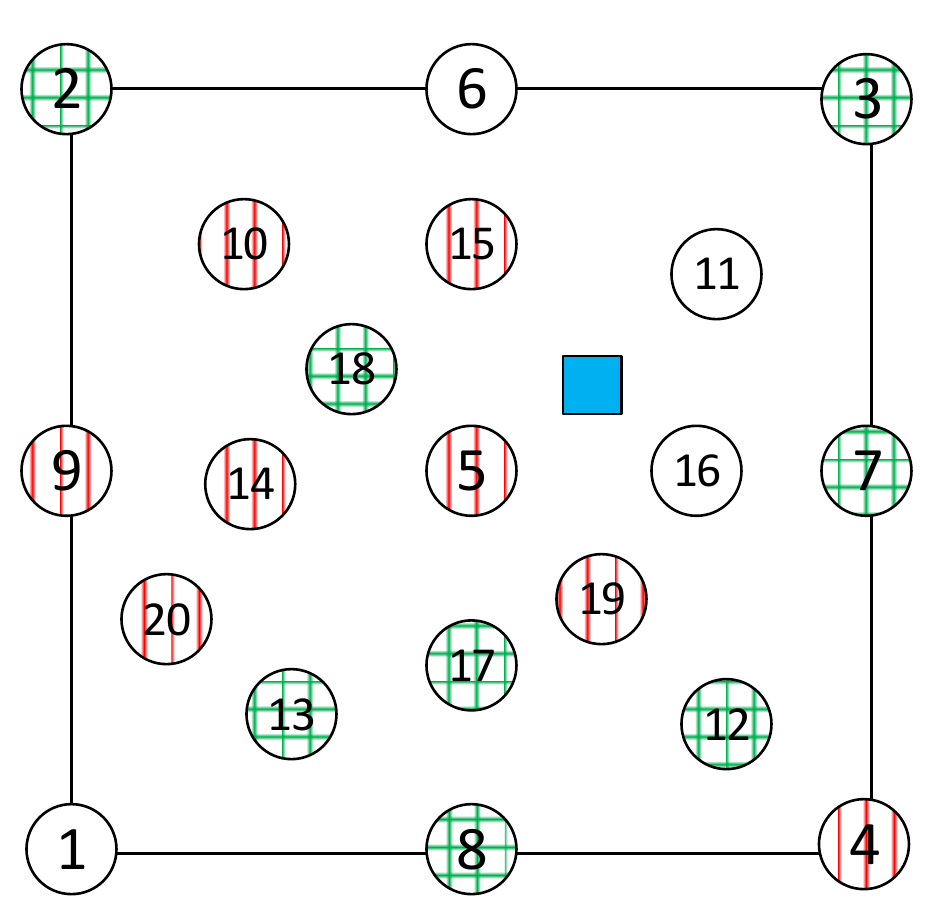} \\ [-0.2cm]
		\end{array}$
		\caption{The topology setup.} 
		\label{fig:Topology}
	\end{center}
\end{figure}

We consider an area with the size of 10$\times$10 km. The positions of charging stations are fixed, while the position of the PEV user will be located randomly in this area. In Fig.~\ref{fig:Topology}, we show a topology to illustrate our simulation in this section. There are 20 charging stations with fixed locations, i.e., circles with numbered labels. The position of the PEV user is located randomly at each simulation and it is illustrated by a blue square in Fig.~\ref{fig:Topology}. There are three prices for charging and discharging, i.e., 0.15, 0.2, 0.25 monetary units (MUs), corresponding to three types of circles, i.e., empty circles, circles with green large grids, and circles with red vertical lines, respectively. For example, if the PEV goes to a charging station which is illustrated by an empty circle, it will pay/receive 0.15 MUs for charging/discharging energy, respectively. The amount of energy to charge/discharge the PEV battery is 60kWh, and the energy consumption is 200Wh per km for traveling. 

\begin{figure}[h]
	\begin{center}
		$\begin{array}{cc}
		\epsfxsize=3.0 in \epsffile{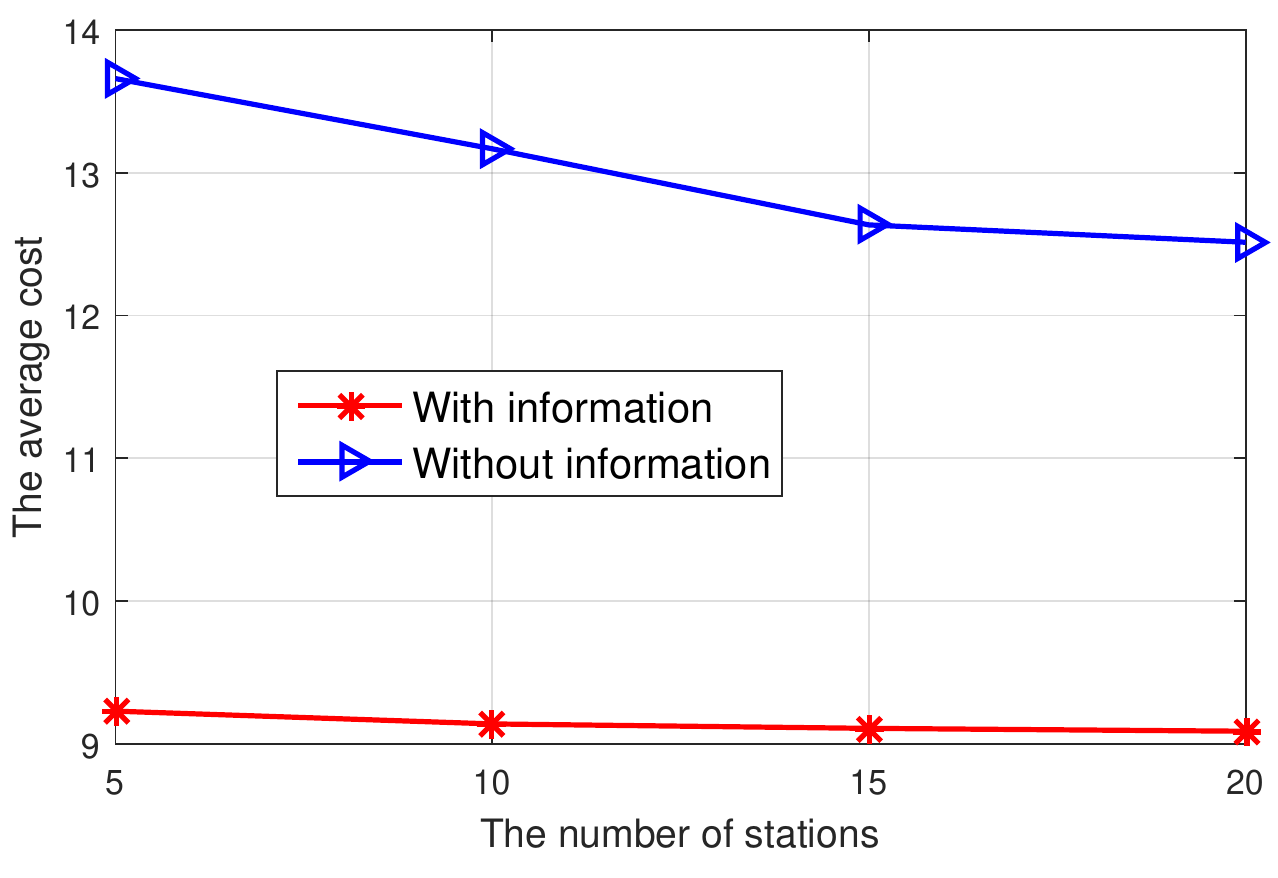}   \\ (a) \\
		\epsfxsize=3.0 in \epsffile{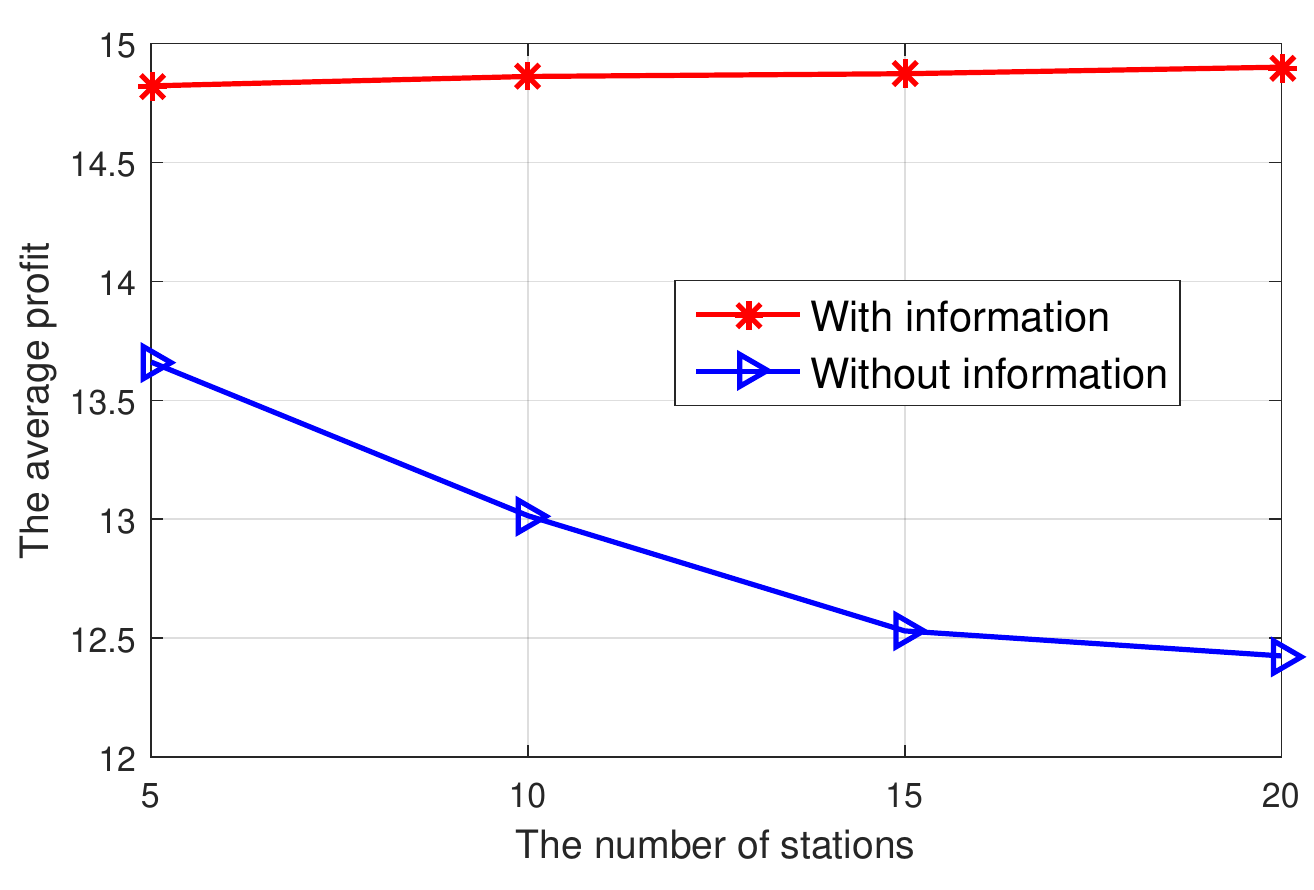}    \\ (b) \\
		\end{array}$
		\caption{(a) Average cost for charging and (b) average profit for discharging.}
		\label{fig:Topology_char/dischar}
	\end{center}
\end{figure}

In Fig.~\ref{fig:Topology_char/dischar}(a) and  Fig.~\ref{fig:Topology_char/dischar}(b), we consider two scenarios, i.e., when the PEV user wants to charge and discharge, respectively. In the case when the infrastructure information is unavailable, the PEV user will find the nearest charging station for charging/discharging, while if the infrastructure information is available, the PEV user will find a charging station which minimizes its cost for charging or maximizes its profit for discharging. The cost of the PEV for charging is equal to the charging cost at the selected station plus the traveling cost, while the discharging profit is equal to the revenue of discharging at the selected station minus the traveling cost. To obtain the average cost as well as the average profit of the PEV user, we perform $50,000$ simulations to calculate the average value. This means that given the topology with a fixed number of stations, the position of the PEV is generated randomly $50,000$ times to find the average value. 

In Fig.~\ref{fig:Topology_char/dischar}, for the case without information, as the number of stations is increased, the average charging cost and discharging profit will be reduced. The reason is that given this topology, when the number of stations is increased, the probability which the PEV user is near the stations with low price will be higher. As a result, both the average charging cost and discharging profit will be decreased in this case (since we set the charging cost and discharging profit to be the same  at a station). However, in the case when the information is available, the average cost/profit slightly increases/decreases as the number of stations increases because the PEV user always can find the best station for charging/discharging to minimize/maximize its cost/profit. In both cases, it is observed that given the infrastructure information, the average cost/profit of the PEV user can be decreased/increased remarkably compared with the case without information. This is from the fact that the PEV has more choices to find a charging station which is not only nearest, but also has the best energy price. This gain is referred to as ``value of information'' which quantifies the benefit of the V2G communication infrastructure. 

However, for the case when the information about V2G infrastructure is unavailable, e.g., due to cyber risks, the PEV user incurs a high cost of charging and gains a low profit from discharging. The cyber insurance can be implemented to ``transfer'' the risks from the PEV user to the insurer. Under the insurance coverage, the PEV user will be guaranteed the best price for charging/discharging at any time. In the following, we will demonstrate the efficiency of using cyber insurance to the V2G system. 

\begin{figure*}[h]
	\begin{center}
		$\begin{array}{cc}
		\epsfxsize=2.5 in \epsffile{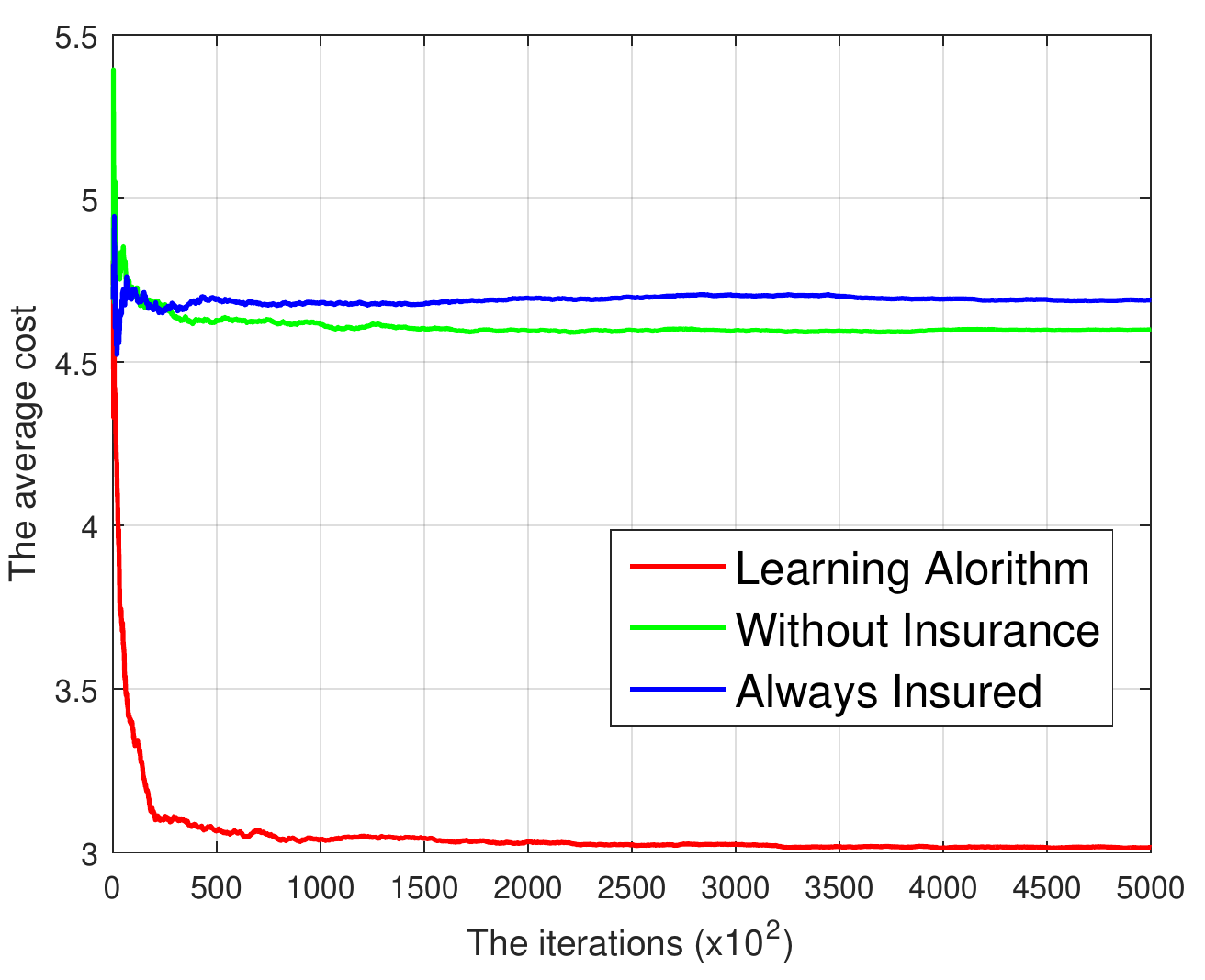}   &
		\epsfxsize=4.0 in \epsffile{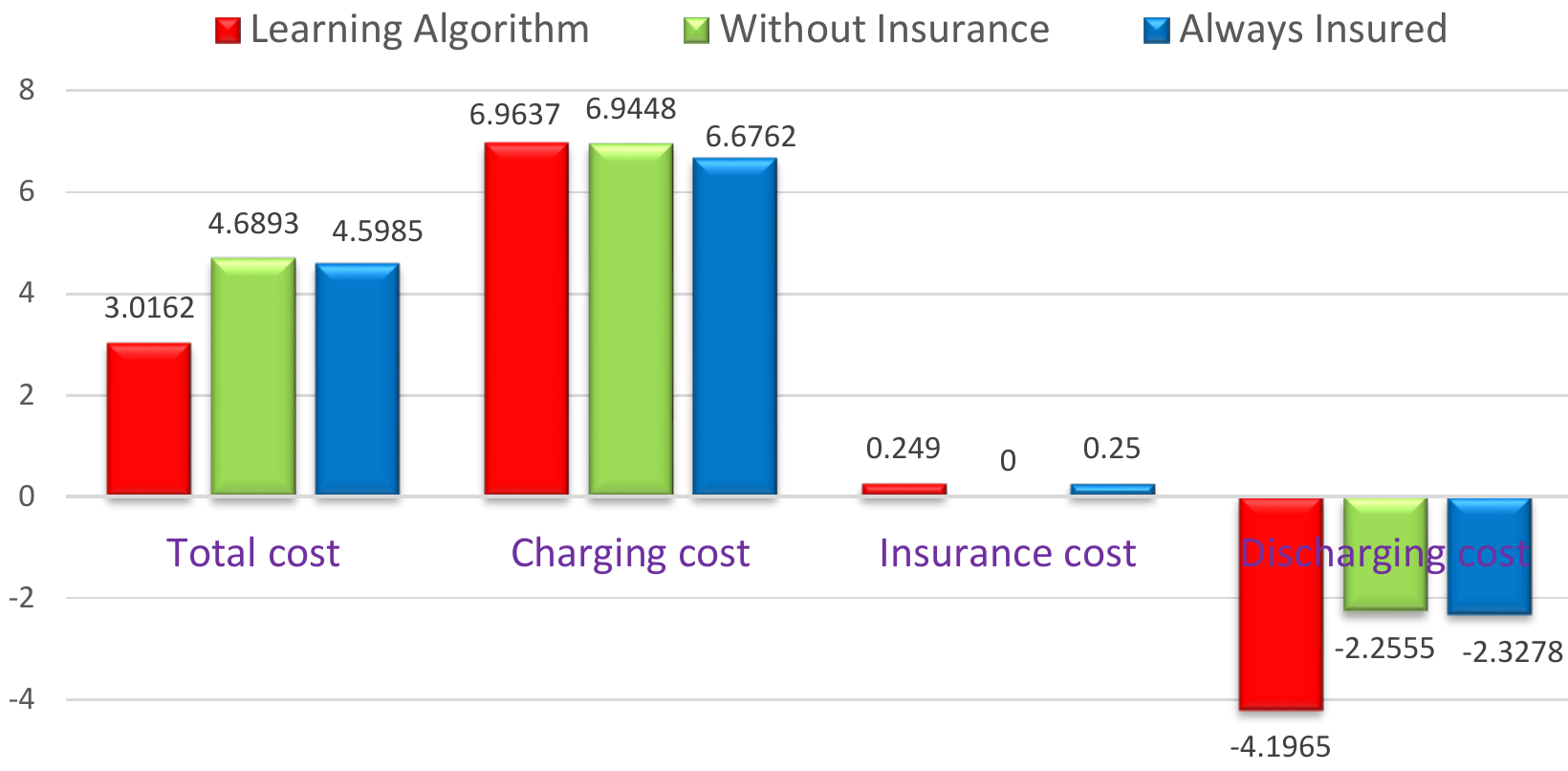}    \\ 
		(a) & (b) 
		\end{array}$
		\caption{(a) The convergence of the learning algorithm and (b) the PEV user's policy.}
		\label{fig:Plot_Convergence_Comparisons}
	\end{center}
\end{figure*}

\begin{figure*}[h]
	\begin{center}
		$\begin{array}{ccc}
		\epsfxsize=2.2 in \epsffile{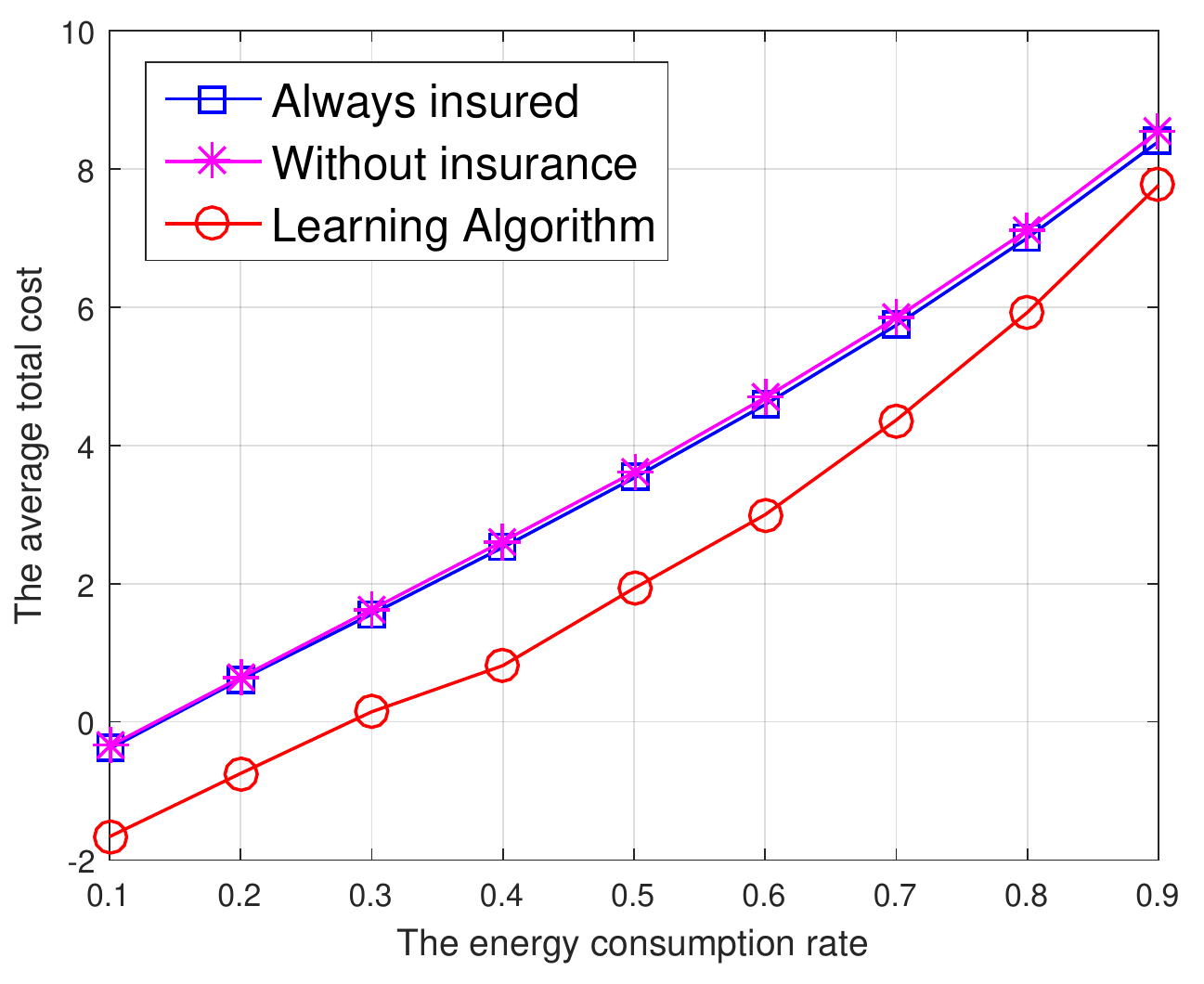}   &
		\epsfxsize=2.2 in \epsffile{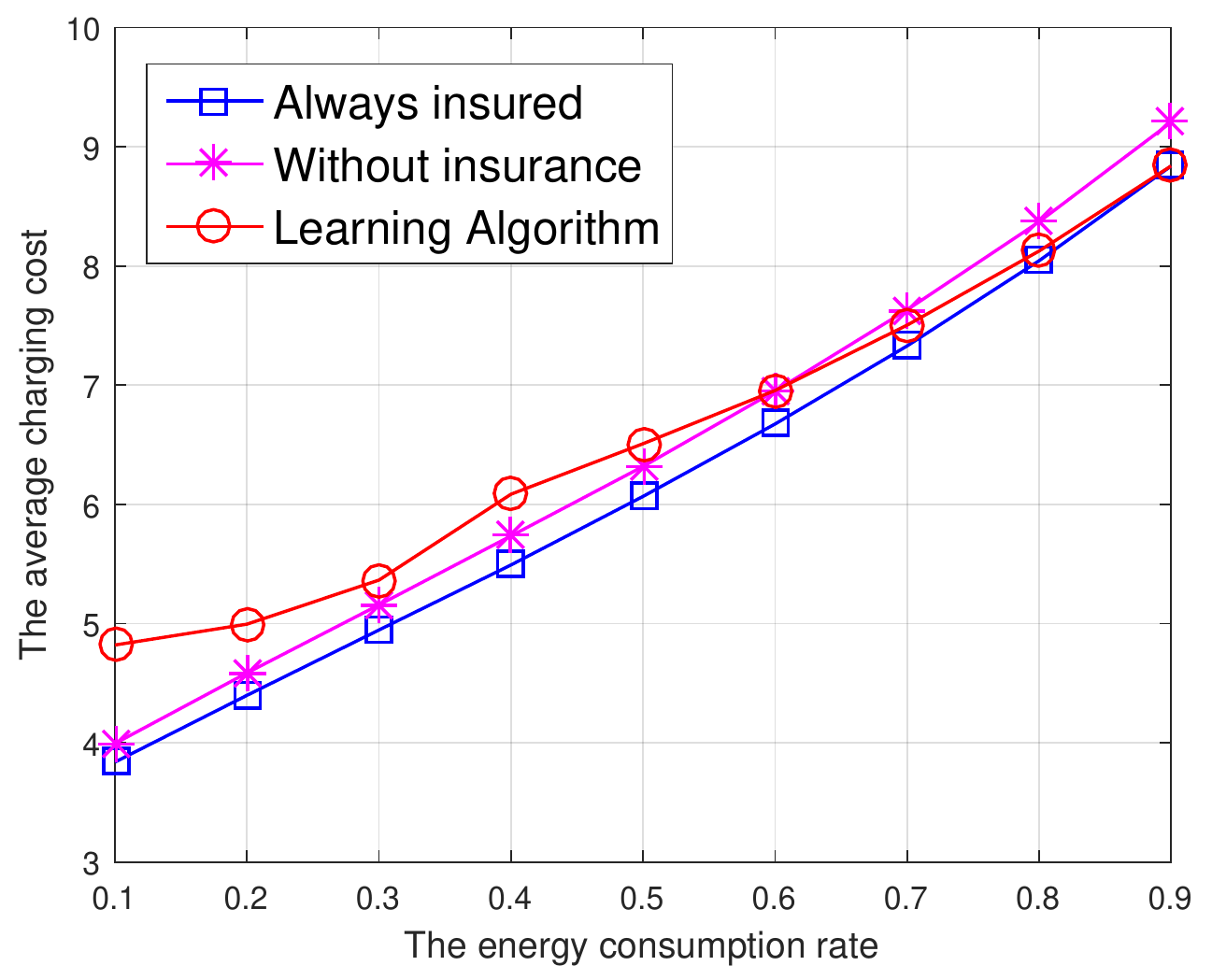}  & 
		\epsfxsize=2.2 in \epsffile{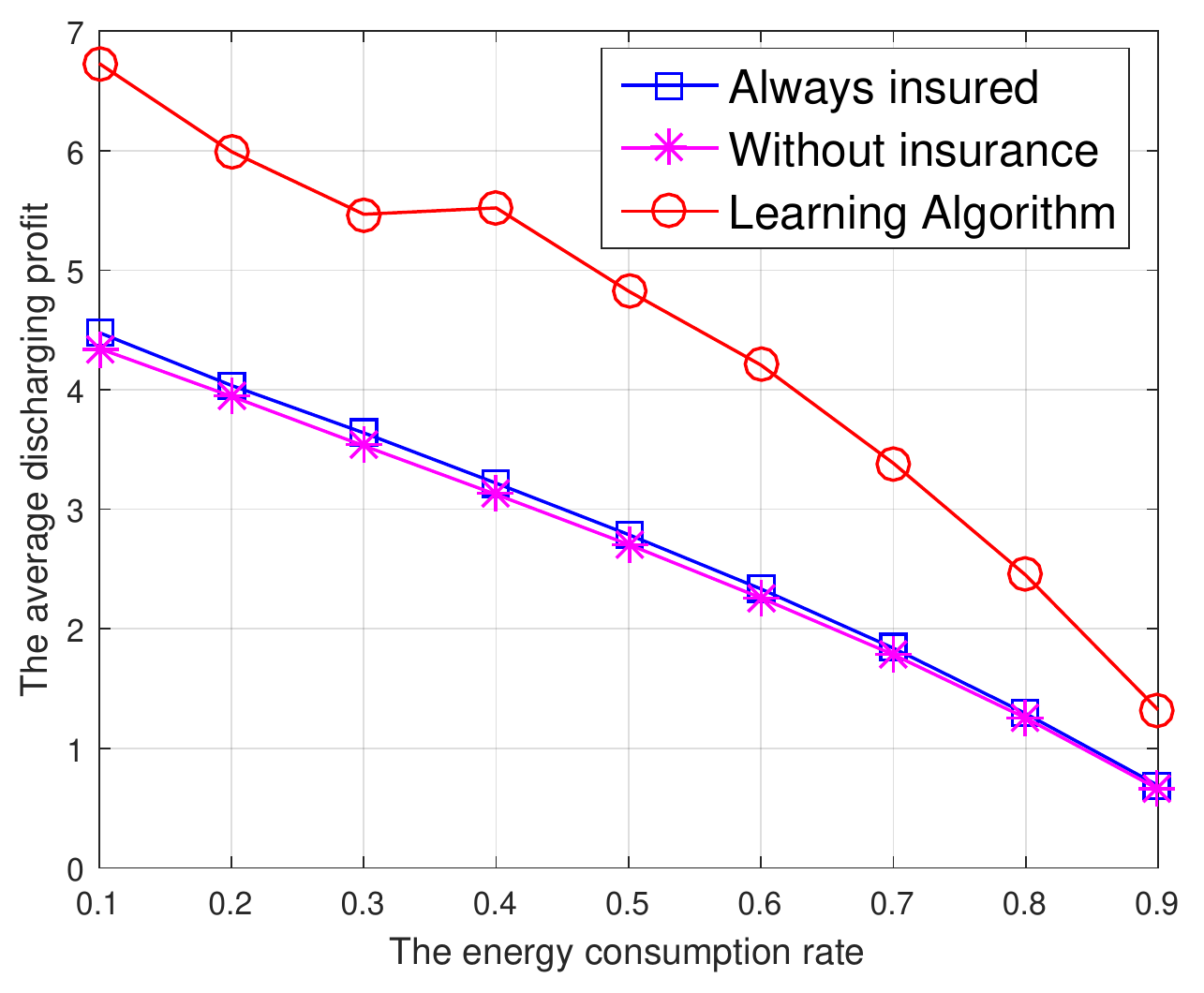} \\
		(a) & (b) & (c) 
		\end{array}$
		\caption{(a) Average total cost, (b) average cost for charging, and (c) average profit of discharging when the energy consumption rate is varied.}
		\label{fig:vary_ene_con}
	\end{center}
\end{figure*}

\subsubsection{Benefits of cyber insurance to the V2G system}

\paragraph{Experiment setup}

The PEV user has a battery with a fixed capacity of 6, i.e., $B=6$, e.g., extremely low, very low, low, moderate, high, and very high levels. There are four periods of time, e.g., morning, afternoon, evening, and night, and there are two insurance status, i.e., insured and not insured. The average charging price when the information is available and unavailable over periods are $[10.5, 10, 9.5, 9]$ and $[14.5, 14, 13.5, 13]$ MUs, respectively. Similarly, the average discharging prices when the information is available and unavailable over periods are $[15.5, 15, 14.5, 14]$ and $[11.5, 11, 10.5, 10]$ MUs, respectively. In the first simulation, i.e., Fig.~\ref{fig:Plot_Convergence_Comparisons}, the energy consumption rate of the PEV user is set at $0.6$, the risk probability is $0.1$, the premium cost is $1$ MU, and the coverage period is $4$ periods (i.e., $\nu=4$). The values of these parameters will be varied later to evaluate the efficiency of the proposed learning algorithm. Here, note that when the information is unavailable and the PEV is under the coverage, the PEV user will be charged at the same price when the information is available. 

In order to evaluate the efficiency of the proposed learning algorithm, i.e., \textbf{Algorithm~\ref{algorithm1}}, we consider two other schemes, i.e., always insured policy (IP) and the policy without insurance (WP). For the IP, the PEV will be always insured, i.e., the PEV will buy insurance every $\nu$-period. For example, if the PEV user buys insurance at time slot $t=1$, then it will buy insurance in time slots $t=1+\nu, 1+2\nu, \ldots$. For both policies, i.e., the IP and the WP, when the energy level is at the lowest level, i.e., $b=1$, the PEV user will always choose action ``charging'' to avoid the heavy cost and prevent energy depletion status. However, when the energy level is higher, i.e., $b\geq2$, the PEV user will select randomly one of three actions, i.e., ``do nothing'', ``charging'', or ``discharging''. For the learning algorithm, the value of the parameter vector $\Theta$ is set at $\bf{0}$, i.e., the PEV user will select 2 actions, i.e., $a_1$ and $a_2$, randomly at the beginning. In other words, at the beginning, the PEV user will select actions ``do nothing'', ``charging'', and ``discharging'' with the same probabilities, i.e., $\frac{1}{3}$, and actions ``buy insurance'' and ``do not buy insurance'' with the same probabilities, i.e., $\frac{1}{2}$. The initial average cost is set at $0$.

\begin{figure*}[h]
	\begin{center}
		$\begin{array}{ccc}
		\epsfxsize=2.2 in \epsffile{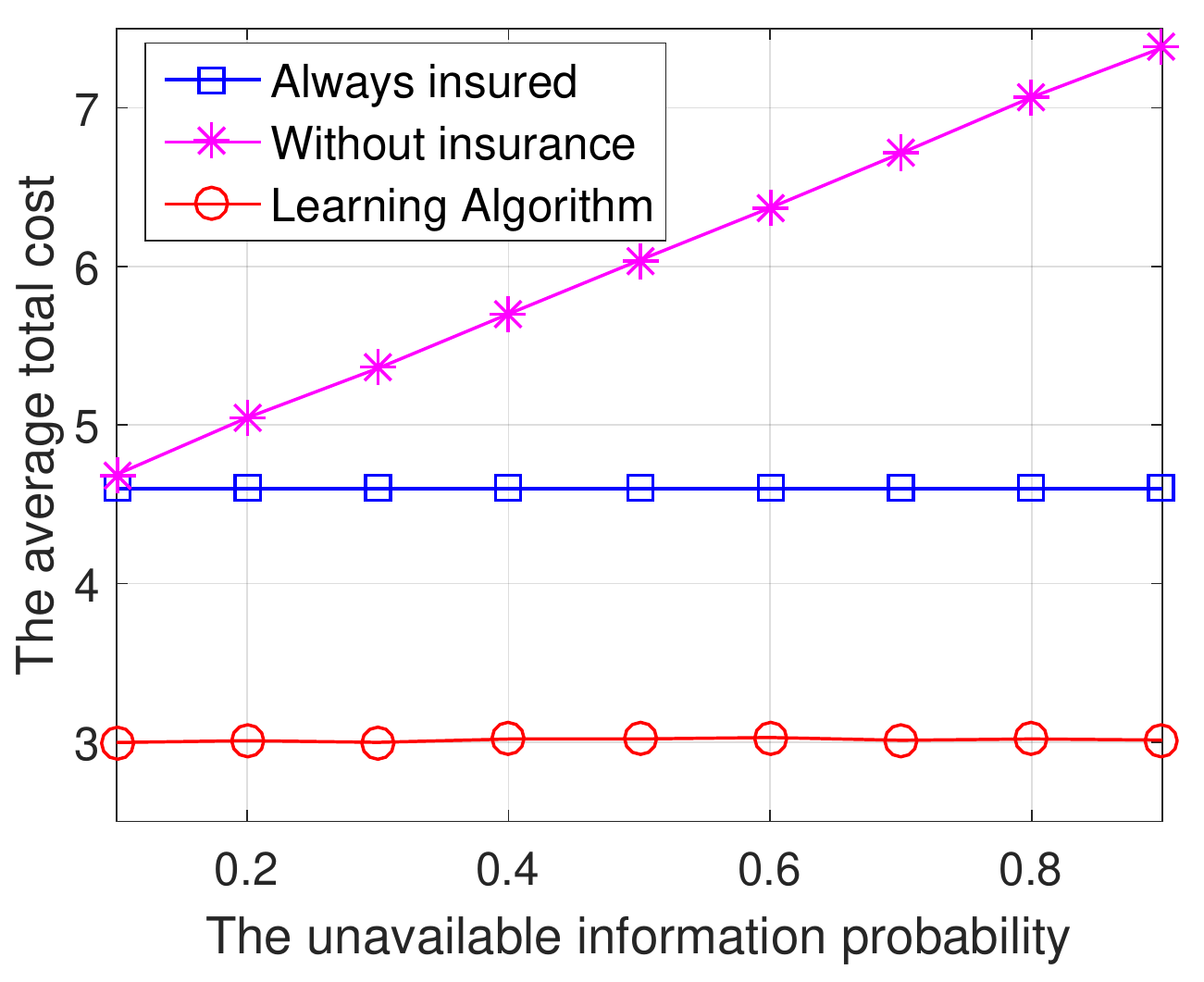}   &
		\epsfxsize=2.2 in \epsffile{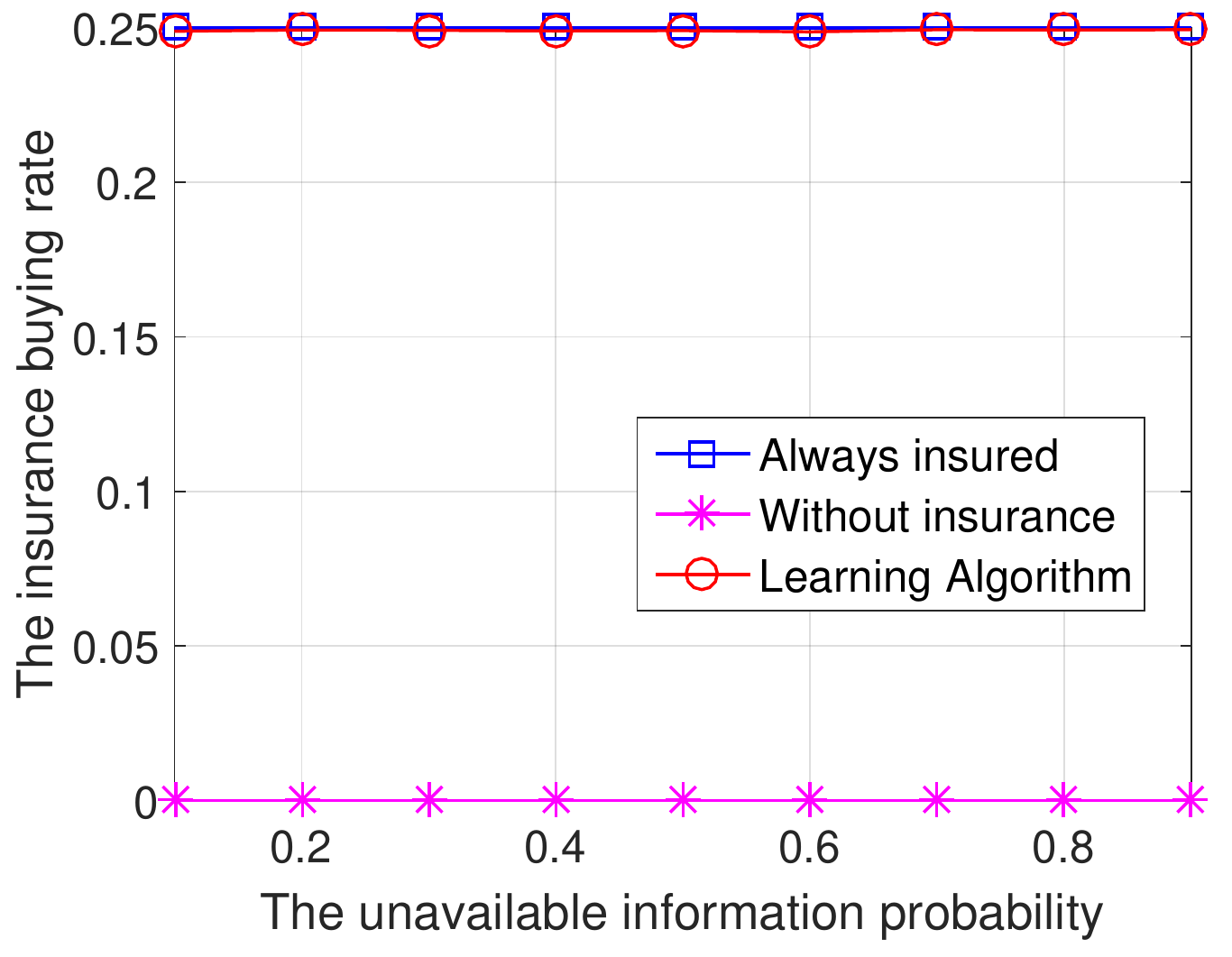}  & 
		\epsfxsize=2.2 in \epsffile{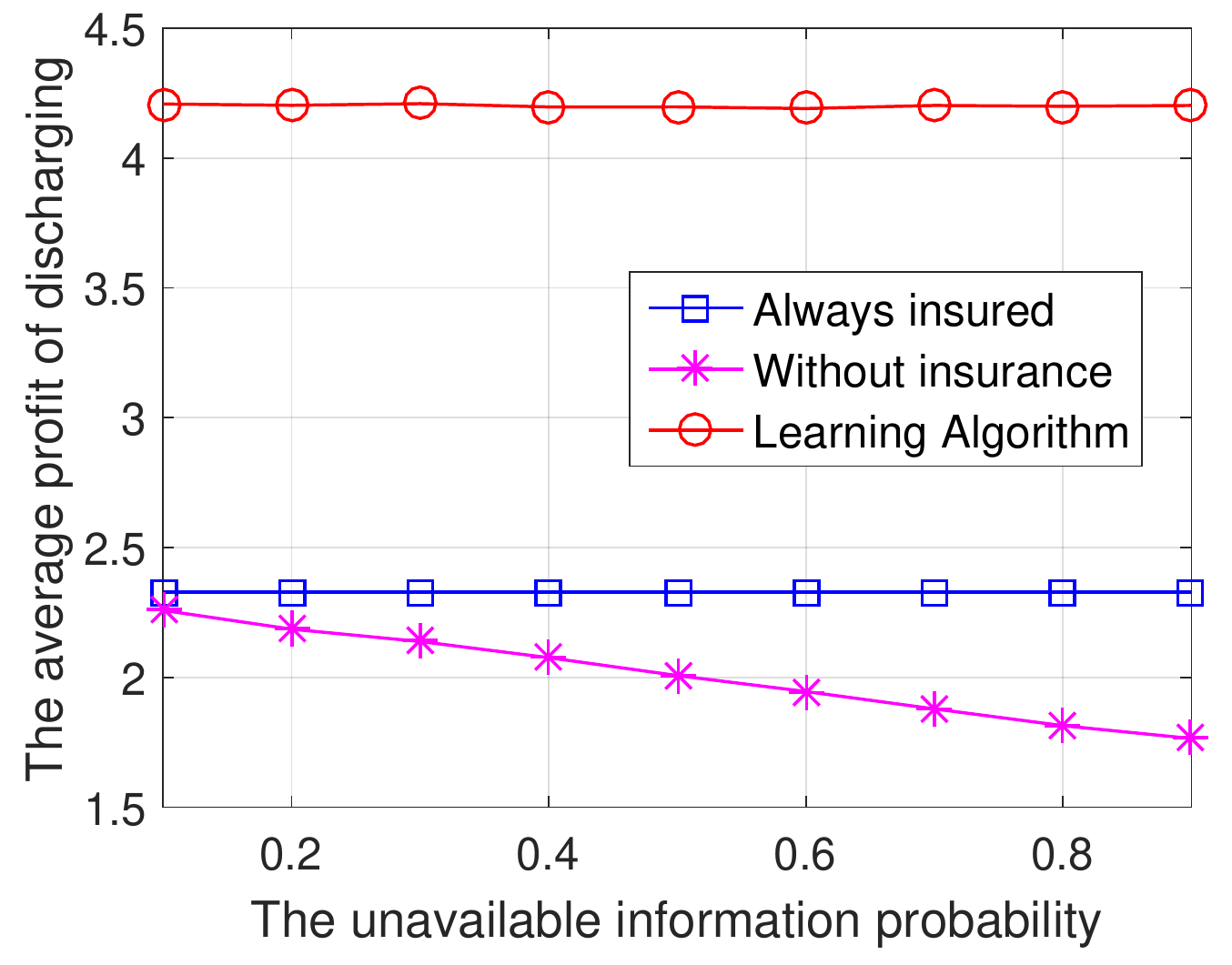} \\
		(a) & (b) & (c) 
		\end{array}$
		\caption{(a) Average total cost, (b) average insurance buying rate and (c) average profit of discharging when the unavailability information probability is varied.}
		\label{fig:vary_info_avail}
	\end{center}
\end{figure*}

\begin{figure*}[h]
	\begin{center}
		$\begin{array}{ccc}
		\epsfxsize=2.2 in \epsffile{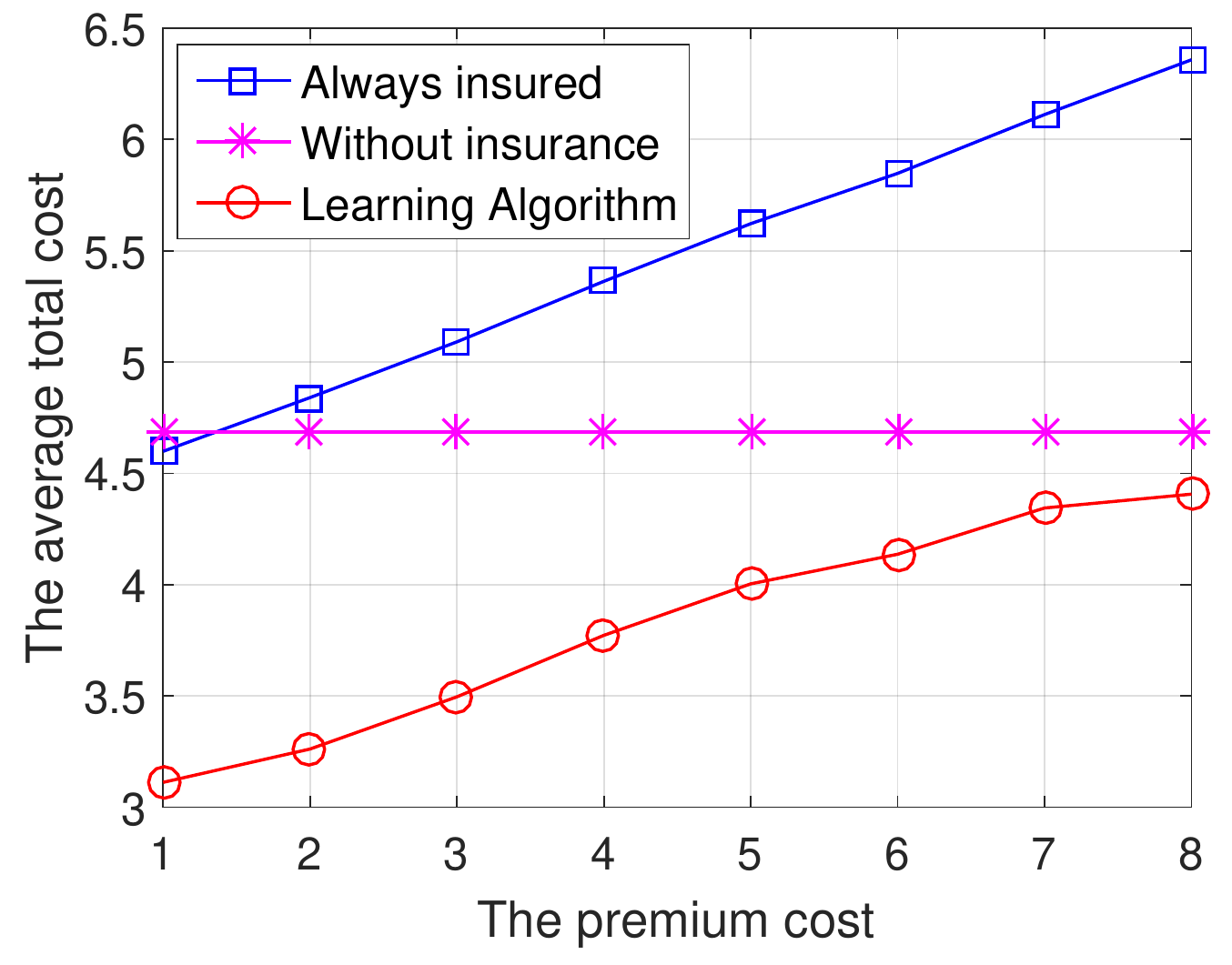}   &
		\epsfxsize=2.2 in \epsffile{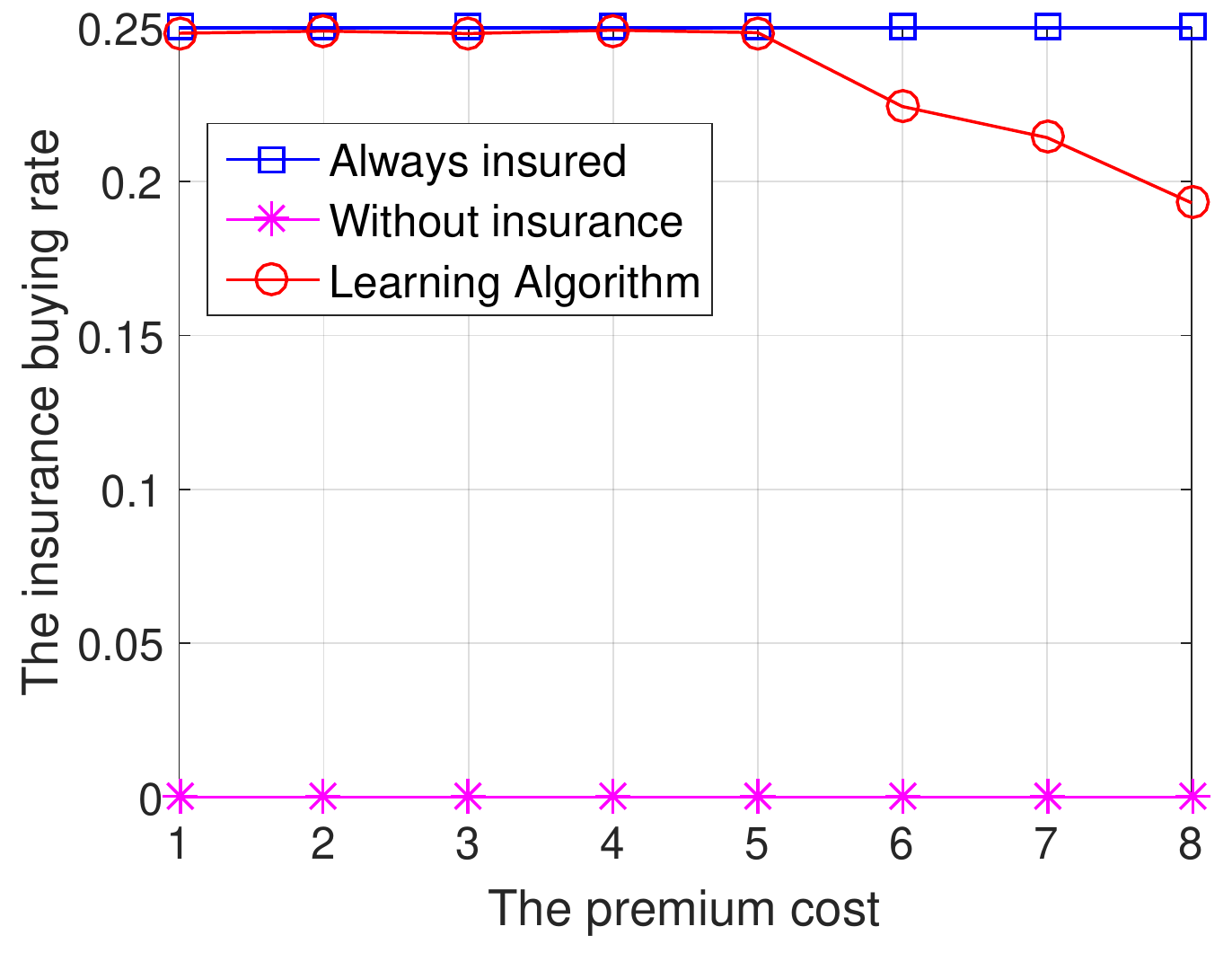}  & 
		\epsfxsize=2.2 in \epsffile{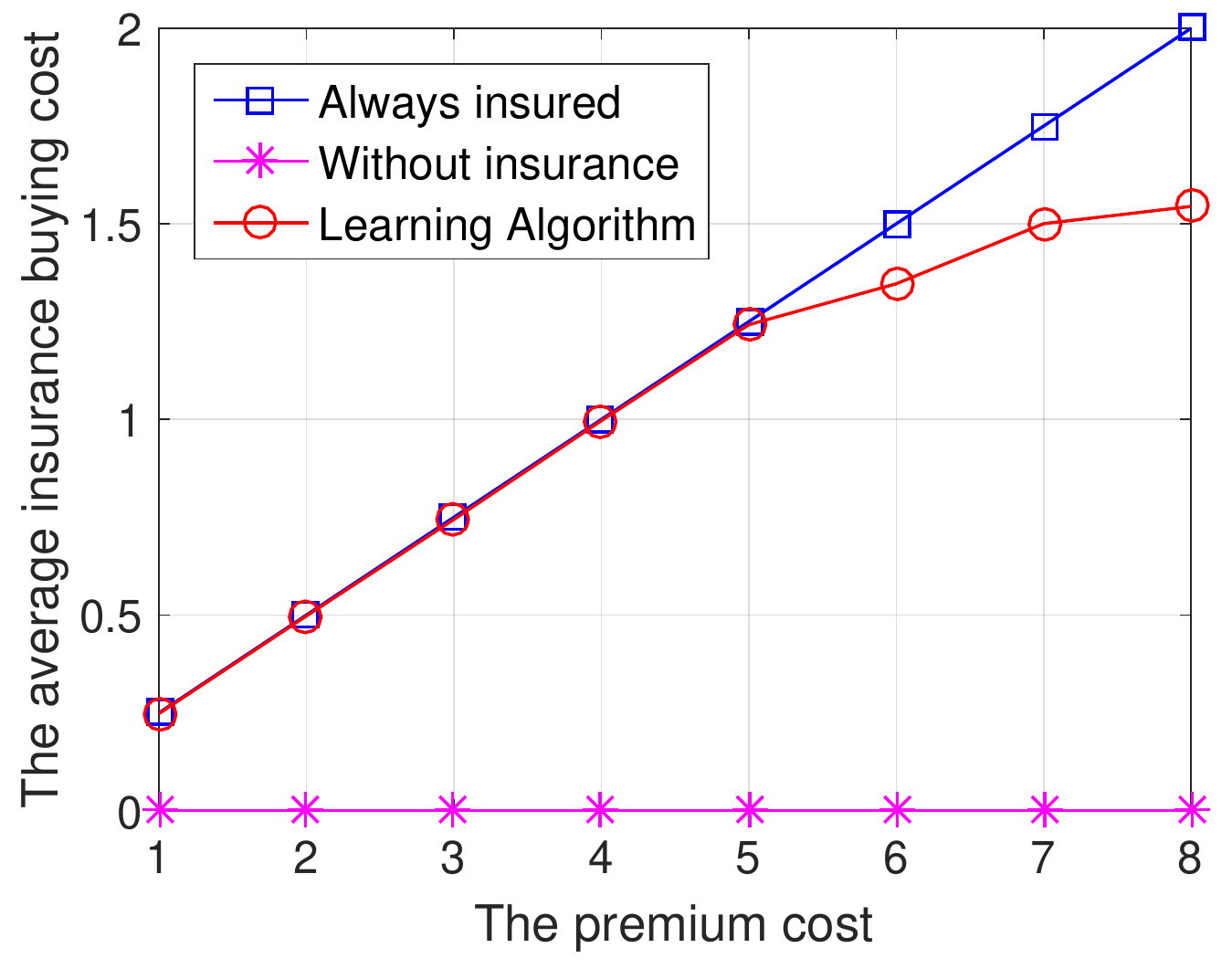} \\
		(a) & (b) & (c) 
		\end{array}$
		\caption{(a) Average total cost, (b) average insurance buying rate, and (c) average insurance buying cost when the premium cost is varied.}
		\label{fig:vary_premium_cost}
	\end{center}
\end{figure*}

\paragraph{Simulation results}

In the simulation, we first show the convergence through the average cost of the proposed learning algorithm. As shown in Fig.~\ref{fig:Plot_Convergence_Comparisons}(a), the average cost of the proposed learning algorithm will converge to approximately 3 when the number of iterations is $10^5$, while the IP and the WP converge to $4.7$ and $4.6$, respectively, after 5$\times 10^4$ iterations. This means that with the proposed learning algorithm, the average cost for the PEV user can be reduced approximately 34.5\% compared with those of the IP and the WP. The efficiency of the proposed learning algorithm can be interpreted through the PEV user's policy in Fig.~\ref{fig:Plot_Convergence_Comparisons}(b). In particular, for the learning algorithm, when the premium cost is set at $m=1$, the PEV user will buy insurance to be insured almost all the time. However, different from the IP, with the learning algorithm, the PEV user can balance among ``charging'', ``discharging'', and `do nothing'' actions to obtain higher profits in discharging and lower cost in charging.

In Fig.~\ref{fig:vary_ene_con}, we vary the energy consumption rate of the PEV user, while other parameters remain unchanged. As the energy consumption rate increases, the average total costs obtained by all policies will be increased as shown in Fig.~\ref{fig:vary_ene_con}(a) because the PEV user needs more energy for its operation. Since the PEV user needs more energy for its operation, the average charging cost is increased as shown in Fig.~\ref{fig:vary_ene_con}(b). Consequently, the discharging process will be reduced which results in a lower discharging profit as shown in Fig.~\ref{fig:vary_ene_con}(c). However, in all cases, the learning algorithm always achieves the best performance in terms of the lowest cost for the PEV user. In Fig.~\ref{fig:vary_ene_con}(a), there is a very interesting point that when the energy consumption is less than 0.3, the average cost of the learning algorithm is less than 0. The reason is that when the energy demand is low, the PEV user still buys energy, i.e., charging, when the energy price is low, and then it will sell, i.e., discharging, when the energy price is high, and thus it can obtain more profits. As a result, the discharging profit is higher than the charging cost (for the case with a low demand of the PEV user), and thus the average total cost is lower than zero. 

We then vary the probability of information unavailability and evaluate the average total cost and the insurance buying rate of the PEV user. Interestingly, at the premium cost $m=1$ MU, when the probability of information unavailability increases from $0.1$ to $0.9$, the average total cost of the WP increases remarkably, while the average total costs	 of the IP and learning algorithm do not change as shown in Fig.~\ref{fig:vary_info_avail}(a). The reason can be explained through the insurance buying policy of the learning algorithm shown in Fig.~\ref{fig:vary_info_avail}(b). In particular, at a low premium cost, i.e., $m=1$ MU for 4 periods, the PEV user will always choose to buy insurance because under the coverage, the PEV user is guaranteed not only the lowest price for charging, but also the highest price for discharging. As a result, the discharging profit obtained by the learning algorithm is always remained at a high level as shown in Fig.~\ref{fig:vary_info_avail}(c), and thus the average cost obtained by the learning algorithm is remained at a low level as shown in Fig.~\ref{fig:vary_info_avail}(a).

Last, we vary the premium cost to evaluate the proposed learning algorithm. In Fig.~\ref{fig:vary_premium_cost}(a), as the premium cost increases, the average total costs of the IP and the learning algorithm increase remarkably. In particular, when the premium cost is higher than 7 MUs, the average total cost obtained by the learning algorithm is close to the average total cost obtained by the WP. The reason is that when the premium cost is too high, the cost to buy insurance will be high (as shown in Fig.~\ref{fig:vary_premium_cost}(c)), diminishing the profit obtained by the insurance, e.g., reducing the charging cost and increasing the discharging profit. Consequently, when the probability of information unavailability is $0.1$, if the premium cost is higher than $5$ MUs, the insurance buying rate obtained by the learning algorithm will be reduced. This analysis is especially important to the insurer to set an appropriate premium to maximize its profits, while still attracting the PEV user in purchasing insurance.

\section{Future Research Directions of Cyber Insurance in V2G systems}
\label{sec:Future Reserach}

In the following, we introduce some future research directions of cyber insurance in V2G systems which not only mitigate risks for PEV users, but also maximize the profit for service providers. 

\subsection{Self-protection Strategy}

Currently, we consider the case when the PEV user has only two decisions, i.e., either to buy or not to buy insurance, to mitigate the risk. However, in practice, the PEV user also can implement self-protection solutions to deal with information unavailability problem, e.g., using a backup energy storage or employing a backup channel to communicate with the V2G system. Thus, the PEV user has to decide to implement its self-protection strategy, buy insurance, or do nothing. In this case, cyber insurance models with self-protection strategy introduced in Section~\ref{subsubsection: Cyber insurance with self-protection} can be adopted to find the optimal policy for the PEV user.

\subsection{Multiple Insurers}

There often exist multiple insurers in practice. Different insurers may have different insurance policies with different charging stations' locations. Furthermore, different PEV users may have different energy demand with different traveling routines. Thus, how to find the best insurer to meet the PEV's requirements and how to set the best insurance price for an insurer given its topology of charging stations are still open questions. To address this problem, stochastic geometry and graph theory can be used. For example, we can model the spatial distribution of the charging stations of an insurer as an $\alpha$-Ginibre point process, and then given the location of a PEV user, we can evaluate the performance for that PEV user in terms of its average overall cost in a similar way as shown in~\cite{Kong2016Extract}.

\subsection{Smart Cyber Insurance Pricing}

In all of the aforementioned scenarios, we assumed that the energy provider is also the service provider, i.e., the insurer, but they can be different entities in general. Consequently, setting a premium is a challenge due to the conflict of interest between the energy provider and the service provider as well as among the service providers. To address this problem, smart pricing strategies can be used. For example, the bundling strategy introduced in~\cite{Niyato2016Smart} can be adopted by multiple service providers to form a coalition and to offer their energy insurance services as a bundle. With bundling, the profit of the service providers can be improved by encouraging PEV users to buy insurance, while the PEV users will be offered more attractive services, e.g., they may have more charging stations to choose from with better insurance prices.

\subsection{Cyber Insurance for V2G systems with Cognitive Radios}

Due to a large number of PEV users, cognitive radios can be considered to be a potential solution to address communication problems for V2G networks~\cite{Khan2016Cognitive}. In cognitive radio networks, P2V users can communicate with V2G infrastructure through primary channels as long as their communication does not cause harmful interference to the primary users~\cite{Hoang_2014_opportunistic}. However, for such networks, the PEV users' communications are uncertain depending on the primary users' demands. Consequently, the information unavailability due to the primary users' communications can cause loss to the PEV users. In this case, cyber insurance can be used as an efficient economic solution to protect the PEV users from risks caused by the information unavailability.

\section{Conclusion}
\label{sec:conclusion}

We have first presented a comprehensive overview on Vehicle-to-Grid (V2G) systems and cyber insurance including basic concepts, general architectures, advantages, and challenges for the development of V2G systems as well as cyber insurance. We have also discussed potential solutions and highlighted some promising future research directions for each topic. Then, we have introduced a new idea of using cyber insurance to mitigate information risks for the V2G system with the aim to mitigate the loss and improve the profit for the Plug-in-Electric Vehicle (PEV) user. In particular, we have demonstrated that without V2G infrastructure information, the average charging cost will be very high, while the average discharging profit will be very low for the PEV user. In addition, we have proposed the learning algorithm which helps the PEV user to make best decisions, i.e., charge/discharge energy and buy/do not buy the insurance, at each time period in an online fashion. Through simulations results, we have showed that the proposed learning algorithm not only minimizes the charging cost, but also maximizes the discharging profit for the PEV user. Furthermore, we have also presented proofs and simulation results to show the convergence of the learning algorithm.

\appendices

\section{The proof of Proposition~\ref{prop:prop_policy_gradient}}
\label{appendix:prop:prop_policy_gradient}

This is to show the gradient of the average cost. In~(\ref{eq: balance equation}), we have $\sum_{s \in \mathcal{S}} \pi_{\Theta}({s})  = 1$, so $\sum_{s \in \mathcal{S}} \nabla \pi_{\Theta}({s})  = 0$. 

Recall that 
\begin{displaymath}
\begin{aligned}
d(s, \Theta)  & = f_c (s, \Theta) - \mathcal{C} (\Theta) + \sum_{s' \in \mathcal{S}} p_b (s'|s,\Psi(\Theta)) d(s', \Theta) ,	\\
\text{and} \phantom{5} & \mathcal{C} (\Theta) = \sum_{s \in \mathcal{S}} \pi_{\Theta}({s}) f_c (s, \Theta).
\end{aligned}
\end{displaymath}

Then, we derive the following results:
\begin{equation}
\begin{aligned}
& \nabla \mathcal{C} (\Theta)  = 	\sum_{s \in \mathcal{S}} \pi_{\Theta}({s}) \nabla f_c (s, \Theta) + \sum_{s \in \mathcal{S}} \nabla \pi_{\Theta}({s}) f_c (s, \Theta)	, \\
= &	\sum_{s \in \mathcal{S}} \pi_{\Theta}({s}) \nabla f_c (s, \Theta) + \sum_{s \in \mathcal{S}} \nabla \pi_{\Theta}({s}) f_c (s, \Theta)	- \\
& \mathcal{C} (\Theta) \sum_{s \in \mathcal{S}} \nabla \pi_{\Theta}({s})	\phantom{5} (\text{since} \phantom{5} \sum_{s \in \mathcal{S}} \nabla \pi_{\Theta}({s}) = 0),	\\
= &	\sum_{s \in \mathcal{S}} \pi_{\Theta}({s}) \nabla f_c (s, \Theta) + \sum_{s \in \mathcal{S}} \nabla \pi_{\Theta}({s})  \big( f_c (s, \Theta) - \mathcal{C} (\Theta) \big) ,	\nonumber	\\
= &	\sum_{s \in \mathcal{S}} \pi_{\Theta}({s}) \nabla f_c (s, \Theta) + \\
& \sum_{s \in \mathcal{S}} \nabla \pi_{\Theta}({s})   \bigg( d(s, \Theta) -  \sum_{s' \in \mathcal{S}} p_b (s'|s,\Psi(\Theta)) d(s', \Theta)  \bigg).
\end{aligned}
\end{equation}

We define
\begin{equation}
\begin{aligned}
\label{appendix:eq:derivation}
&\nabla \Big( \pi_{\Theta}({s})  p_b (s'|s,\Psi(\Theta)) \Big) = \\ 
&\nabla \pi_{\Theta}({s}) p_b (s'|s,\Psi(\Theta)) + \pi_{\Theta}({s}) \nabla p_b (s'|s,\Psi(\Theta))	,
\end{aligned}
\end{equation}
and from~(\ref{eq: balance equation}), $\sum_{s \in \mathcal{S}} \pi_{\Theta}({s}) p_b(s'|s,\Psi(\Theta)) = \pi_{\Theta}({s'})$. Then, we have the derivations as given in~(\ref{xxxx1}) (next page).

\begin{figure*}[!t]
	\normalsize
	\begin{equation}
	\label{xxxx1}
	\begin{aligned}
	\nabla C(\Theta)	& = \sum_{s \in \mathcal{S}} \pi_{\Theta}(s) \nabla\mathcal{C} (\Theta)  + \sum_{s \in \mathcal{S}} \nabla \pi_{\Theta}(s) 	\bigg( d(s, \Theta) 	-  \sum_{s' \in \mathcal{S}} p_b (s'|s,\Psi(\Theta)) d(s', \Theta)	 \bigg)	\\
	& =  \sum_{s \in \mathcal{S}} \pi_{\Theta}(s) \nabla\mathcal{C} (\Theta) +  \sum_{s \in \mathcal{S}} \nabla  \pi_{\Theta}(s) d(s, \Theta) + \sum_{s,s' \in \mathcal{S}} \bigg( \pi_{\Theta}(s) \nabla p_b (s'|s,\Psi(\Theta)) - \nabla\Big( \pi_{\Theta}(s) \nabla p_b (s'|s,\Psi(\Theta)) \Big) \bigg) d(s', \Theta) \\
	& = \sum_{s \in \mathcal{S}} \pi_{\Theta}(s) \nabla\mathcal{C} (\Theta)  +  \sum_{s \in \mathcal{S}} \nabla  \pi_{\Theta}(s) d(s, \Theta) + \sum_{s,s' \in \mathcal{S}} \pi_{\Theta}(s) \nabla p_b (s'|s,\Psi(\Theta)) d(s', \Theta) - \\
	& \phantom{10} \sum_{s' \in \mathcal{S}} \nabla \Big( \sum_{s \in \mathcal{S}} \pi_{\Theta}(s)  p_b (s'|s,\Psi(\Theta))  \Big) d(s', \Theta)		\\
	& = \sum_{s \in \mathcal{S}} \pi_{\Theta}(s) \nabla\mathcal{C} (\Theta)  +  \sum_{s \in \mathcal{S}} \nabla  \pi_{\Theta}(s) d(s, \Theta)	+  \sum_{s,s' \in \mathcal{S}} \pi_{\Theta}(s) \nabla p_b (s'|s,\Psi(\Theta)) d(s', \Theta)  - \sum_{s' \in \mathcal{S}} \nabla \pi_{\Theta}(s') d(s', \Theta)	\\
	& = \sum_{s \in \mathcal{S}} \pi_{\Theta}(s)	\bigg( \nabla \mathcal{C} (\Theta)  + \sum_{s' \in \mathcal{S}} \nabla p_b (s'|s,\Psi(\Theta)) d(s', \Theta)	\bigg)
	\end{aligned}
	\end{equation}
	\hrulefill
	\vspace*{-4pt}
\end{figure*}

The proof now is completed.

\section{The proof of Proposition~\ref{prop2}}
\label{appendix:prop2}
We will prove the convergence of the Algorithm~\ref{algorithm0}. The update equations of Algorithm~\ref{algorithm0} can be rewritten in the specific form as in~(\ref{xxxx_2}) (next page).
\begin{figure*}[!t]
	\normalsize
	\begin{equation}
	\label{xxxx_2}
	\begin{aligned}
	&\Theta_{m+1} = \Theta_{m} + \rho_{m} \left( \sum_{t'=t_{m}}^{t_{m+1}-1} \Big( \sum_{t=t'}^{t_{m+1}-1}( f_c(s_t,a_t) - \widetilde{\psi}_{m}) \Big) \frac{\nabla \mu_{\Theta_{m}}(s_{t'},a_{t'})}{\mu_{\Theta_{m}}(s_{t'},a_{t'})} \right), \\
	&\widetilde{\psi}_{m+1} = \widetilde{\psi}_{m} + \kappa\rho_{m}\sum_{t'=t_{m}}^{t_{m+1}-1}( f_c (s_{t}, a_{t}) - \widetilde{\psi}_{m})
	\end{aligned}
	\end{equation}
	\vspace*{-15pt}
	\hrulefill
\end{figure*}

We define the vector $\mathbf{r}^{k_m}=\left[	\begin{array}{cc}	\Theta_m	&	\widetilde{\psi}_{m}	\end{array}	\right]^\top$, then~(\ref{xxxx_2}) becomes
\begin{equation}
\label{xxxx_3}
\mathbf{r}^{k_{m+1}} = \mathbf{r}^{k_m} + \rho_{m} \mathbf{H}_m,
\end{equation}
where 
\begin{equation}
\begin{aligned}
&\mathbf{H}_m \!	 = \\
& \!	\left[\!	\begin{array}{c}
\sum_{t'=t_{m}}^{t_{m+1}-1} \Big( \sum_{t=t'}^{t_{m+1}-1}( f_c (s_{t}, a_{t}) - \widetilde{\psi}_{m}) \Big) \frac{\nabla \mu_{\Theta_{m}}(s_{t'},a_{t'})}{\mu_{\Theta_{m}}(s_{t'},a_{t'})}	\\
\kappa \sum_{t'=t_{m}}^{t_{m+1}-1}( f_c (s_{t}, a_{t}) - \widetilde{\psi}_{m})
\end{array}	\! \right].
\end{aligned}
\end{equation}

Let $\mathscr{F} = \{ \Theta_0, \widetilde{\psi}_{0}, s_0,s_1,\ldots,s_m  \}$ be the history of the Algorithm~\ref{algorithm0}. Then from Proposition 2 in~\cite{Marbach2001}, we have
\begin{equation}
\begin{aligned}
\mathbb{E}[\mathbf{H}_m|\mathscr{F}_m]	& \!=\! \mathbf{h}_m	\!=\! \left[	\begin{array}{c}
\mathbb{E}_{\Theta}[T] \nabla \mathcal{C} (\Theta) + \mathscr{V}(\Theta) \big( \mathcal{C} (\Theta) - \widetilde{\psi}(\Theta) \big)	\\
\kappa \mathbb{E}_{\Theta}[T] \big( \mathcal{C} (\Theta) - \widetilde{\psi}(\Theta)	\big)
\end{array}	\right],
\end{aligned}
\end{equation}
where 
\begin{displaymath}
\mathscr{V}(\Theta)=\mathbb{E}_{\Theta}\Bigg[ \sum_{t'=t_{m+1}}^{t_{m+1}-1} \big( t_{m+1} - t'\big) \frac{\nabla \mu_{\Theta_{m}}(s_{t'},a_{t'})}{\mu_{\Theta_{m}}(s_{t'},a_{t'})} \Bigg].
\end{displaymath}

Consequently, (\ref{xxxx_3}) has the following form
\begin{equation}
\mathbf{r}^{k_{m+1}} = \mathbf{r}^{k_m} + \rho_{m} \mathbf{h}_m + \varepsilon_m,
\end{equation}
where $\varepsilon_m = \rho (\mathbf{H}_m - \mathbf{h}_m)$ and note that $\mathbb{E}[\varepsilon_m|\mathscr{F}_m]=0$. Since $\varepsilon_m$ and $\rho_m$ converge to zero almost surely, along with the fact that $\mathbf{h}_{m}$ is bounded, we have
\begin{equation}
\lim_{m \rightarrow \infty} (\mathbf{r}^{k_{m+1}}-\mathbf{r}^{k_{m}}) = 0.
\end{equation}

After that, based on Lemma 11 in~\cite{Marbach2001}, it is proved that $\psi(\Theta)$ and $\widetilde{\psi}(\Theta)$ converge to a common limit. This means the parameter vector $\Theta$ can be represented in the following way
\begin{equation}
\label{eq:appendix_standard_form}
\Theta_{m+1} = \Theta_{m} + \rho_m \mathbb{E}_{\Theta_m}[T] \big(\nabla \mathcal{C} (\Theta_m) + e_m \big) + \epsilon_m,
\end{equation}
where $e_m$ is an error term that converges to zero and $\epsilon_m$ is a summable sequence.~(\ref{eq:appendix_standard_form}) is known as the gradient method with diminishing errors~\cite{Bertsekas1999,Borkar2008}. Therefore, following the same way in~\cite{Bertsekas1999,Borkar2008}, we can prove that $\nabla \mathcal{C} (\Theta_m)$ converges to $0$, i.e., $\nabla_{\Theta} \mathcal{C} (\Theta_{\infty})=0$.

\section*{Acknowledgements}

This work was supported in part by Singapore MOE Tier 1 (RG18/13 and RG33/12) and MOE Tier 2 (MOE2014-T2-2-015 ARC4/15 and MOE2013-T2-2-070 ARC16/14).


\clearpage
\begin{IEEEbiography}[{\includegraphics[width=1in,height=1.25in,clip,keepaspectratio]{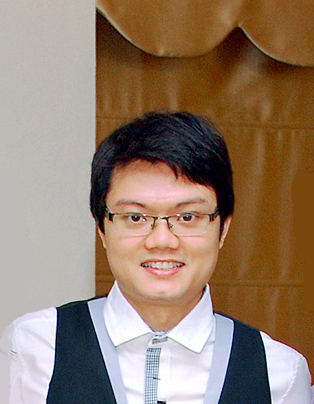}}] {Dinh Thai Hoang} (M'16) received the Ph.D. degree in 2016 from School of Computer Science and Engineering, Nanyang Technological University, Singapore, where he is currently a Research Fellow. His research interests include optimization problems and game theory for wireless communication networks and mobile cloud computing.
\end{IEEEbiography}
\vspace{-13cm}

\begin{IEEEbiography}[{\includegraphics[width=1in,height=1.25in,clip,keepaspectratio]{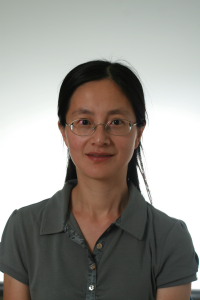}}] {Ping Wang} (M'08-SM'15) received the Ph.D. degree in electrical engineering from University of Waterloo, Canada, in 2008. Currently she is an Associate Professor in the School of Computer Science and Engineering, Nanyang Technological University, Singapore. Her current research interests include resource allocation in wireless networks, cloud computing, and smart grid. She was a corecipient of the Best Paper Award from IEEE Wireless Communications and Networking Conference (WCNC) 2012 and IEEE International Conference on Communications (ICC) 2007.
\end{IEEEbiography}
\vspace{-13cm}

\begin{IEEEbiography}[{\includegraphics[width=1in,height=1.25in,clip, keepaspectratio]{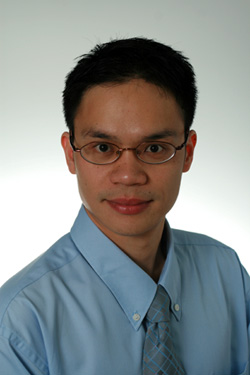}}]{Dusit Niyato} (M'09-SM'15-F'17) is currently an Associate Professor in the School of Computer Science and Engineering, at Nanyang Technological University, Singapore. He received B.Eng. from King Mongkut's Institute of Technology Ladkrabang (KMITL), Thailand in 1999 and Ph.D. in Electrical and Computer Engineering from the University of Manitoba, Canada in 2008. His research interests are in the area of energy harvesting for wireless communication, Internet of Things (IoT) and sensor networks.
\end{IEEEbiography}

\newpage
\begin{IEEEbiography} [{\includegraphics[width=1in,height=1.25in,clip,keepaspectratio]{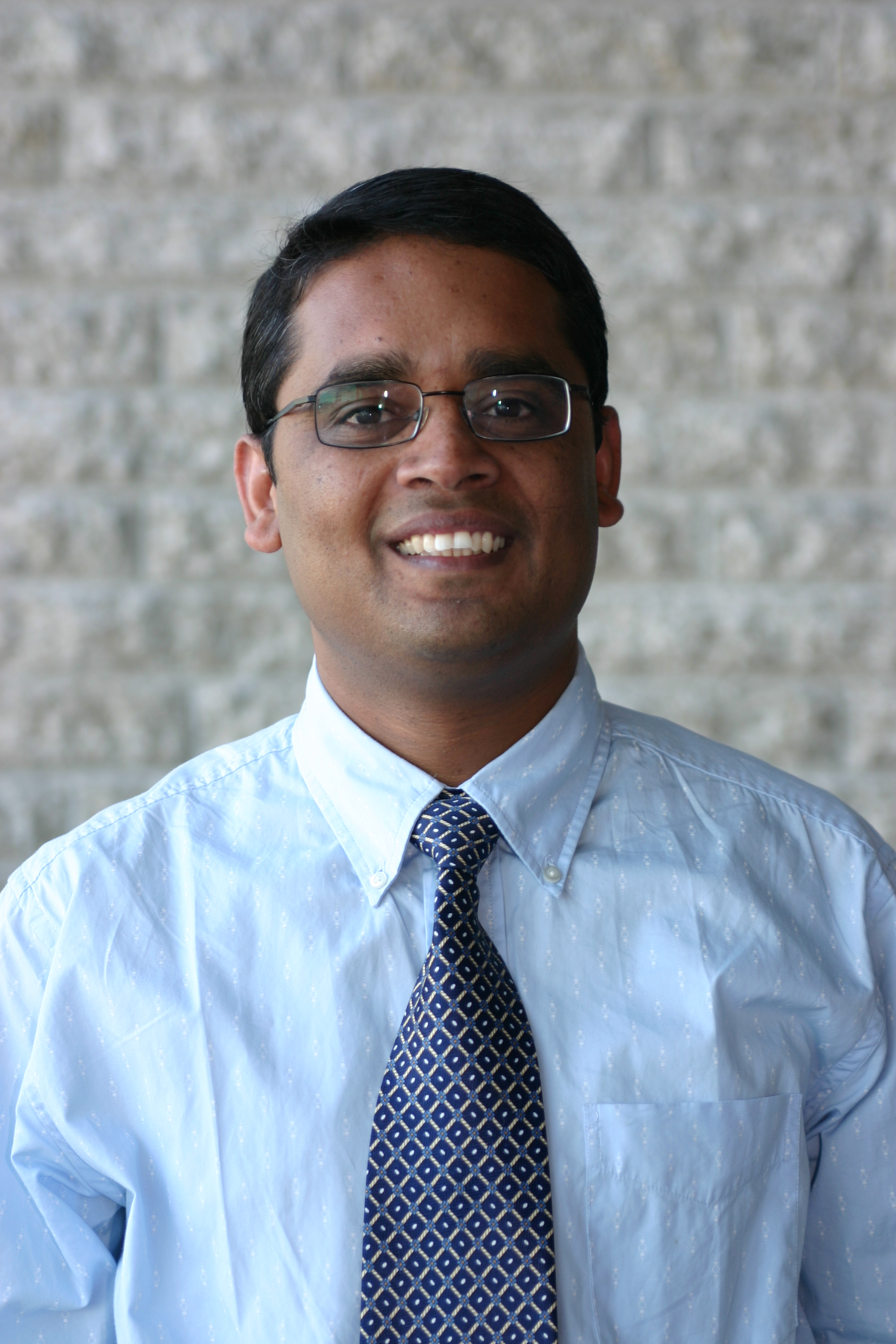}}]
{Ekram Hossain} (F'15) is a Professor (since March 2010) in the Department of Electrical and Computer Engineering at University of Manitoba, Winnipeg, Canada. He is a Member (Class of 2016) of the College of the Royal Society of Canada. He received his Ph.D. in Electrical Engineering from University of Victoria, Canada, in 2001. He was elevated to an IEEE Fellow ``for spectrum management and  resource allocation in  cognitive and cellular radio networks".  Dr. Hossain's current research interests include design, analysis, and optimization of wireless/mobile communications networks, cognitive radio systems, and network economics.  He has authored/edited several books in these areas (http://home.cc.umanitoba.ca/$\sim$hossaina). He  serves as the Editor-in-Chief for the {\em IEEE Communications Surveys and Tutorials}  and an Editor for  {\em IEEE Wireless Communications}.  Also, he is a member of the IEEE Press Editorial Board. Previously, he served as the Area Editor for the {\em IEEE Transactions on Wireless Communications} in the area of  ``Resource Management and Multiple Access'' from 2009-2011, an Editor for the {\em IEEE Transactions on Mobile Computing} from 2007-2012, and an Editor for the {\em IEEE Journal on Selected Areas in Communications - Cognitive Radio Series} from 2011-2014. Dr. Hossain has won several research awards including the IEEE Vehicular Technology Conference (VTC 2016 - Fall) Best Student Paper Award as a co-author,  IEEE Communications Society Transmission, Access, and Optical Systems (TAOS) Technical Committee's Best Paper Award in IEEE Globecom 2015, University of Manitoba Merit Award in 2010, 2014, and 2015 (for Research and Scholarly Activities), the 2011 IEEE Communications Society Fred Ellersick Prize Paper Award, and the IEEE Wireless Communications and Networking Conference 2012 (WCNC'12) Best Paper Award. He was elevated to an IEEE Fellow ``for spectrum management and  resource allocation in  cognitive and cellular radio networks".  Dr. Hossain was a Distinguished Lecturer of the IEEE Communications Society (2012-2015). Currently he is a  Distinguished Lecturer of the IEEE Vehicular Technology Society. He is a registered Professional Engineer in the province of Manitoba, Canada.
\end{IEEEbiography}


\begin{thebibliography}{100}
\bibliographystyle{IEEEtranS}

\bibitem{Gordon2003AFrame}
L.~A.~Gordon, M.~P.~Loeb, and T.~Sohail, ``A framework for using insurance for cyber-risk management,'' \emph{Communications of the ACM}, vol. 46, no. 3, pp. 81--85, Mar. 2003.

\bibitem{Wei2014GT-CFS} 
C.~Wei, Z.~M.~Fadlullah, N~Kato and A.~Takeuchi, ``GT-CFS: A game theoretic coalition formulation strategy for reducing power loss in micro grids,'' \emph{IEEE Transactions on Parallel and Distributed Systems}, vol. 25, no. 9, pp. 2307--2317, Sept. 2014.

\bibitem{Tan2016Pareto} 
X.~Tan, Y.~Wu, and D.~H.~K.~Tsang, ``Pareto optimal operation of distributed battery energy storage systems for energy arbitrage under dynamic pricing,'' \emph{IEEE Transactions on Parallel and Distributed Systems}, vol. 27, no. 7, pp. 2103--2115, July 2016.

\bibitem{Kempton2005Vehicle}
W.~Kempton and J.~Tomic, ``Vehicle-to-grid power fundamentals: Calculating capacity and net revenue,'' \emph{Journal of Power Sources}, vol. 144, no. 1, pp. 268-279, Jun. 2005. 

\bibitem{Guille2009A conceptual}
C.~Guille and G.~Gross, ``A conceptual framework for the vehicle-to-grid (V2G) implementation,'' \emph{Energy Policy}, vol. 37, no. 11, pp. 4379--4390, Nov. 2009.

\bibitem{Villalobos2014Plug}
J.~García-Villalobos, I.~Zamora, J.~I.~San Martín, F.~J.~Asensio, and V.~Aperribay, ``Plug-in electric vehicles in electric distribution networks: A review of smart charging approaches,'' \emph{Renewable and Sustainable Energy Reviews}, vol. 38, pp. 717-731, Oct. 2014.

\bibitem{Yang2011P2}
Z. Yang, S. Yu, W. Lou, and C. Liu, ``$P^2$: Privacy-preserving communication and precise reward architecture for V2G networks in smart grid,'' \emph{IEEE Transactions on Smart Grid}, vol. 2, no. 4, pp. 697--706, Dec. 2011.

\bibitem{Islam2014Integrating}
F.~R.~Islam and H.~R.~Pota, ``Integrating smart PHEVs in future smart grid,'' in \emph{Renewable Energy Integration}, pp. 239-258, Springer Singapore, 2014. 

\bibitem{Battery_Type}
BU-1003: Electric Vehicle (EV), \url{http://batteryuniversity.com/learn/article/electric_vehicle_ev}

\bibitem{Lam2011ZigBee}
K.~L.~Lam, K.~T.~Ko, H.~Y.~Tung, H.~C.~Tung, K.~F.~Tsang, and L.~L.~ Lai, ``ZigBee electric vehicle charging system,'' in \emph{IEEE International Conference on Consumer Electronics}, pp. 507-508, Las Vegas, US, Jan. 2011. 

\bibitem{Steffen2010Near}
R.~Steffen, J.~Preibinger, T.~Schollermann, A.~Muller, and I.~Schnabel, ``Near field communication (NFC) in an automotive environment,'' in \emph{International Workshop on  Near Field Communication}, pp. 15-20, Grimaldi Forum, Monaco, Apr. 2010.

\bibitem{Conti2011B4V2G}
M.~Conti, D.~Fedeli, and M.~Virgulti, ``B4V2G: Bluetooth for electric vehicle to smart grid connection,'' in \emph{Proceedings of the Ninth Workshop on Intelligent Solutions in Embedded Systems}, pp. 13--18, Regensburg, Germany, Jul. 2011. 

\bibitem{Anbagi2016WAVE}
I.~Al-Anbagi and H.~T.~Mouftah, ``WAVE 4 V2G: Wireless access in vehicular environments for vehicle-to-grid applications,'' \emph{Vehicular Communications}, pp. 31--42, vol. 3, Jan. 2016.

\bibitem{Msadaa2010A comparative}
I.~C.~Msadaam, P.~Cataldi, and F.~Filali, ``A comparative study between 802.11p and mobile WiMAX-based V2I communication networks,'' in \emph{International Conference on Next Generation Mobile Applications, Services and Technologies}, pp. 186--191, Amman , Jordan, July, 2010. 

\bibitem{Jatav2014WiMAX}
V.~K.~Jatav and V.~Singh, ``Mobile WiMAX network security threats and solutions: A survey,'' in \emph{International Conference on Computer and Communication Technology}, pp. 135--140, Allahabad, India, Sept. 2014.

\bibitem{Liu2014Role}
H.~Liu, H.~Ning, Y.~Zhang, Q.~Xiong, and L.~T.~ Yang, ``Role-dependent privacy preservation for secure V2G networks in the smart grid,'' \emph{IEEE Transactions on Information Forensics and Security}, vol. 9, no. 2, pp. 208-219, Feb. 2014.

\bibitem{Shuaib2016CognitiveRadio}
K.~Shuaib, E.~Barka, N.~A.~Hussien, M.~Abdel-Hafez, and M.~Alahmad, ``Cognitive radio for smart grid with security considerations,'' \emph{Computers}, vol. 5, no. 2, pp. 7, Apr. 2016.

\bibitem{Xu2005The feasibility}
W.~Xu, W.~Trappe, Y.~Zhang, T.~Wood, ``The feasibility of launching and detecting jamming attacks in wireless networks,'' in \emph{Proceedings of the 6th ACM International Symposium on Mobile ad hoc Networking and Computing}, pp. 25-28, Urbana-Champaign, IL, USA, May 2005.

\bibitem{Xu2004Channel}
W.~Xu, T.~Wood, W.~Trappe, Y.~Zhang, ``Channel surfing and spatial retreats: Defenses against wireless denial of service,'' in \emph{Proceedings of the 3rd ACM Workshop on Wireless Security, Philadelphia}, pp. 80-89, PA, USA, Sept. 2004.

\bibitem{Hoang2015Performance}
D.~T.~Hoang, D.~Niyato, P.~Wang, and D.~I.~Kim, ``Performance analysis of wireless energy harvesting cognitive radio networks under smart jamming attacks,'' \emph{IEEE Transactions on Cognitive Communications and Networking}, vol. 1, no. 2, pp. 200-216, June 2015.

\bibitem{Pelechrinis2011Denial}
K.~Pelechrinis, M.~Iliofotou, and S.~V.~Krishnamurthy, ``Denial of service attacks in wireless networks: The case of jammers,'' \emph{IEEE Communications Surveys \& Tutorials}, vol. 13, no. 2, pp. 245-257, May 2011.

\bibitem{Pal2014Will}
R.~Pal, L.~Golubchik, K.~Psounis, and P.~Hui, ``Will cyber-insurance improve network security? A market analysis,'' in \emph{IEEE Conference on Computer Communications}, pp. 235--243, Toronto, Canada, May, 2014.

\bibitem{Majuca2005_The}
R.~P.~Majuca, W.~Yurcik, and J.~P.~Kesan, ``The evolution of cyber insurance,'' \emph{Information Systems Frontier}, 2005.

\bibitem{online_forbe}
\url{http://www.forbes.com/sites/stevemorgan/2015/12/24/cyber-insurance-market-storm-forecast-2-5-billion-in-2015-projected-to-reach-7-5-billion-by-2020/#2613f69e3ffe}

\bibitem{Kesan2004The}
J.~Kesan, R.~Majuca, and W.~Yurcik. ``The Economic Case for Cyberinsurance'' (July 2004). University of Illinois Law and Economics Working Papers. Working Paper 2.

\bibitem{Lelarge2009Economic}
M.~Lelarge and J.~Bolot, ``Economic incentives to increase security in the Internet: The case for insurance,'' in \emph{IEEE Conference on Computer Communications}, pp. 1494--1502, Rio de Janeiro, Brazil, Apr. 2009. 

\bibitem{Poletti1998_First}
T.~Poletti, ``First-ever insurance against hackers,'' June 1998, available http://goo.gl/SSGArI on 13/07/2015.

\bibitem{Srinidhi2015Allocation}
B.~Srinidhi, J.~Yan, and G.~K.~Tayi, ``Allocation of resources to cyber-security: The effect of misalignment of interest between managers and investors,'' in \emph{Decision Support Systems}, vol. 75, pp. 49--62, May 2015. 

\bibitem{Ishikawa2016AStudy}
T.~Ishikawa and K.~Sakurai, ``A Study of Security Management with Cyber Insurance,'' in \emph{Proceedings of the 10th International Conference on Ubiquitous Information Management and Communication}, pp. 68--73, 2016. 

\bibitem{Laszka2015Should}
A.~Laszka and J.~Grossklags, ``Should cyber-insurance providers invest in software security,'' in \emph{European Symposium on Research in Computer Security}, pp. 483-502, Springer International Publishing, 2015. 

\bibitem{Saini2011Utility}
D.~K.~Saini, I.~Azad, N.~B.~Raut, and L.~A.~Hadimani, ``Utility implementation for cyber risk insurance modeling,'' in \emph{Proceedings of the World Congress on Engineering}, vol. 1, 2011. 

\bibitem{Adeleke2011Cyber}
I.~A.~Adeleke, A.~Ibiwoye, F.~F.~Olowokudejo, ``Cyber risk exposure and prospects for cyber insurance,'' \emph{International Journal of Management and Business Research}, vol. 1, no. 4, pp. 221--230, Aug 2011. 

\bibitem{Cebula2010ATaxonomy}
J.~J.~Cebula and L.~R.~Young, ``A taxonomy of operational cyber security risks,'' Technical Note CMU/SEI-2010-TN-028, CERT Carnegie Mellon University, Dec. 2010.

\bibitem{Pal2010Analyzing}
R.~Pal and L.~Golubchik, ``Analyzing self-defense investments in Internet security under cyber-insurance coverage,'' in \emph{International Conference on Distributed Computing System}, pp. 339--347, Genoa, Italy, June 2010.

\bibitem{Elnagdy2016Cyber}
S.~A.~Elnagdy, M.~Qiu, and K.~Gai, ``Cyber incident classification using ontology-based knowledge representation for cybersecurity insurance in financial industry,'' in \emph{International Conference on Cyber Security and Cloud Computing}, pp. 301--306, Beijing, China, June 2016.

\bibitem{Toregas2014Insurance}
C.~Toregas and N.~Zahn, ``Insurance for cyber attacks: The issues of setting premiums in context,'' Technical Report, The George Washington University, Jan. 2014. 

\bibitem{Pandey2014Applicability} 
P.~Pandey and E.~A.~Snekkenes, ``Applicability of prediction markets in information security risk management,'' in \emph{International Workshop on Database and Expert Systems Applications}, pp. 296-300, Munich, Germany, Sep. 2014. 

\bibitem{Wolfers2006Prediction}
J.~Wolfers and E.~Zitzewitz, ``Prediction markets in theory and practice,'' \emph{National Bureau of Economic Research}, Working Paper 12083, Mar. 2006.

\bibitem{Pandey2015Anovel}
P.~Pandey and S.~D.~Haes, ``A novel financial instrument to incentivize investments in information security controls and mitigate residual risk,'' in \emph{nternational Conference on Emerging Security Information, Systems and Technologies}, pp. 23--28, Venice, Italy, Aug. 2015. 

\bibitem{Pal2013On}
R.~Pal, L.~Golubchik, K.~Psounis, and P.~Hui, ``On a way to improve cyber-insurer profits when a security vendor becomes the cyber-insurer,'' in \emph{IEEE IFIP Networking Conference}, pp. 1--9, New York, USA, May, 2013.

\bibitem{Gollier2004Book}
C.~Gollier. \emph{The Economics of Risk and Time}. MIT Press, 2004.

\bibitem{Mossin1968Aspects}
J.~Mossin, ``Aspects of rational insurance purchasing,'' \emph{Journal of Political Economy}, vol. 76, no. 4, pp. 553--568, Aug. 1968.

\bibitem{Bolot2008Cyber}
J.~Bolot and M.~Lelarge, ``Cyber insurance as an incentive for Internet security,'' in \emph{Seventh Workshop on Economics of Information Security}, pp. 1--19, Hanover, US, Jun. 2008.

\bibitem{Ehrlich1972Market}
I.~Ehrlich and G.~S.~Becker, ``Market insurance, self-insurance, and self-protection, \emph{The Journal of Political Economy}, vol. 80, no. 4, pp. 623--648, 1972.

\bibitem{Kunreuther2003Interdependet}
H.~Kunreuther and G.~Heal, ``Interdependent security: the case of identical agents,'' \emph{Journal of Risk and Uncertainty}, vol. 26, no. 2, pp. 231--249, 2003.

\bibitem{Pal2012Cyber}
R.~Pal, ``Cyber-insurance for cyber-security a solution to the information asymmetry problem,'' \emph{SIAM Annual Meeting}, May, 2012. 

\bibitem{Pal2011CyberAegis}
R.~Pal, G.~Leana, and P.~Konstantinos, ``Aegis: A novel cyber-insurance model,'' in \emph{International Conference on Decision and Game Theory for Security}, pp. 131--150, Nov. 2011. 

\bibitem{Herath2011Copula}
H.~Herath, and H.~Tejaswini, ``Copula-based actuarial model for pricing cyber-insurance policies,'' \emph{Insurance Markets and Companies: Analyses and Actuarial Computations}, vol. 2, no. 1, pp. 7--20, Feb. 2011.

\bibitem{chaisiri2015}
S. Chaisiri, R. Ko, and D. Niyato, ``A joint optimization approach to security-as-a-service allocation and cyber insurance management," in {\em Proceedings of IEEE International Conference on Trust, Security and Privacy in Computing and Communications (IEEE TrustCom)}, Helsinki, Finland, 20-22 August, 2015.

\bibitem{Gai2016ANovel}
K.~Gai, M.~Qiu, and S.~A.~Elnagdy, ``A novel secure big data cyber incident analytics framework for cloud-based cybersecurity insurance,'' in \emph{International Conference on Big Data Security on Cloud}, pp. 171--176, New York, US, Apr. 2016.

\bibitem{Marbach2001}
P.~Marbach, and J.~N.~Tsitsiklis, ``Simulation-based optimization of Markov reward processes,'' in \emph{IEEE Transactions on Automatic Control}, vol. 46, pp. 191--209, Feb. 2001. 

\bibitem{Olivier2007}
O.~Buffet, A.~Dutech, and F.~Charpillet, ``Shaping multi-agent systems with gradient reinforcement learning,'' \emph{Journal of Autonomous Agents and Multi-Agent System}, vol. 15, pp. 197--220, Jan. 2007.

\bibitem{Baxter2001}
J.~Baxter, P.~L.~Barlett, L.~Weaver, ``Experiments with infinite-horizon, policy-gradient estimation,'' \emph{Journal of Artificial Intelligence Research}, vol. 15, pp. 351--381, Nov. 2001.

\bibitem{Bertsekas_1995_Nonlinearprogramming}
Dimitri P. Bertsekas, {\em Nonlinear Programming}. Athena Scientific, Belmont, MA, 1995.

\bibitem{Kong2016Extract}
H-B.~Kong, I.~Flin, P.~Wang, D.~Niyato, and N.~Privault, ``Extract performance analysis of ambient RF energy harvesting wireless sensor networks with Ginibre point process,'' \emph{IEEE Journal on Selected Areas in Communications}, Oct. 2016. 

\bibitem{Niyato2016Smart}
D.~Niyato, D.~T.~Hoang, N.~C.~Luong, P.~Wang, D.~I.~Kim, and Z.~Han, ``Smart data pricing models for Internet-of-Things (IoT): A bundling strategy approach,'' \emph{IEEE Network}, vol. 30, no. 2, pp. 18-25, March 2016.

\bibitem{Khan2016Cognitive}
A.~A.~Khan, M.~H.~Rehmani, and M.~Reisslein, ``Cognitive radio for smart grids: Survey of architectures, spectrum sensing mechanisms, and networking protocols,'' \emph{IEEE Communications Surveys \& Tutorials}, vol. 18, no. 1, pp. 860-998, Oct. 2016.

\bibitem{Hoang_2014_opportunistic}
D.~T.~Hoang, D.~Niyato, P.~Wang, D.~I.~Kim, ``Opportunistic channel access and RF energy harvesting in cognitive radio networks,'' \emph{IEEE Journal of Selected Areas in Communications}, vol. 32, no. 11, November 2014.

\bibitem{Bertsekas1999}
D.~P.~Bertsekas, and J.~N.~Tsitsiklis, ``Gradient convergence in gradient methods with errors,'' in \emph{SIAM Journal on Optimization}, vol. 10, issue 3, pp. 627--642, 1999. 

\bibitem{Borkar2008}
V.~S.~Borkar, {\em Stochastic Approximation: A Dynamic Systems Viewpoint}. Cambridge University Press, 2008. 





\end{thebibliography}
\end{document}